\documentclass[aps,prb,twocolumn,groupedaddress,amsmath,amssymb,amsfonts,graphics]{revtex4}

\DeclareMathOperator{\cycles}{\#}
\DeclareMathOperator{\tr}{tr}

\usepackage{graphicx}
\usepackage{bm}
\usepackage{bbold}

\begin{document}

\title{Some formal results for the valence bond basis}

\author{K.\ S.\ D.\ Beach}
\affiliation{Department of Physics, Boston University, 590 Commonwealth Avenue, Boston, MA 02215}
\author{Anders W.\ Sandvik}
\affiliation{Department of Physics, Boston University, 590 Commonwealth Avenue, Boston, MA 02215}

\date{May 3, 2006}

\begin{abstract}
In a system with an even number of SU(2) spins, there is an overcomplete set of 
states---consisting of all possible pairings of the spins into valence bonds---that spans
the $S=0$ Hilbert subspace. Operator expectation values in this basis are related to the 
properties of the closed loops that are formed by the overlap of valence bond states.
We construct a generating function for spin correlation functions of arbitrary order
and show that all nonvanishing contributions arise from configurations that are
topologically irreducible. We derive explicit formulas for the correlation functions
at second, fourth, and sixth order. We then extend the valence bond basis to include triplet bonds
and discuss how to compute properties that are related to operators acting outside the singlet sector.
These results are relevant to analytical calculations and to numerical valence bond
simulations using quantum Monte Carlo, variational wavefunctions, or exact diagonalization.

\end{abstract}

\maketitle

\section{Introduction \label{SEC:Introduction}}

\begin{figure}
\begin{center}
\includegraphics{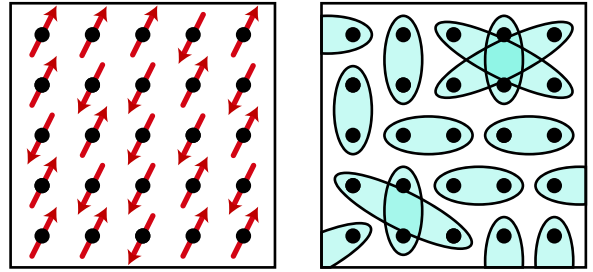}
\end{center}
\caption{ \label{FIG:valence-Sz}
The low-energy singlet sector of quantum antiferromagnets can be described
in either the $S^z$ or valence bond basis.
}
\end{figure}

The traditional way to represent the quantum states of a system of $S=1/2$ spins is to 
introduce a basis of $S^z$ eigenstates. Each state corresponds to a particular assignment 
of ``up'' or ``down'' to each spin in the lattice. This basis is complete and orthonormal. 
Another useful basis, dating back to Rumer and Pauling in the 1930s,\cite{Rumer32,Pauling33} 
is the so-called \emph{valence bond basis}, in which the states of the system are 
represented by partitions of the spins into pairs forming singlets. A valence bond (VB) state, 
\begin{equation} \label{EQ:vbstate}
\lvert v \rangle = \prod_{ij} [i,j],
\end{equation}
is a product of singlets,
\begin{equation}
[i,j] = \frac{1}{\sqrt{2}}(\lvert \uparrow_i \downarrow_j \rangle - \lvert \downarrow_i \uparrow_j \rangle),
\end{equation}
in which each site label appears only once. Pictorially, a VB state can be represented as 
a system of hard-core dimers, where each lattice site (spin) belongs to exactly one dimer,
as illustrated in Fig.~\ref{FIG:valence-Sz}. The VB basis spans the singlet sector of the 
Hilbert space, but is massively overcomplete. It is also highly nonorthogonal, having the 
unusual property that every two states in the  basis have nonzero overlap.

The first important work in this basis was the calculation of the ground-state energy per 
spin of the infinite quantum Heisenberg chain by Hulth\'{e}n~\cite{Hulthen38} (building 
on the earlier work of Bethe~\cite{Bethe31}). Hulth\'{e}n also determined the eigenvalues 
and eigenstates of small finite chains of up to ten sites using the subset of  
``noncrossing''~\cite{Rumer32} valence bond states (which form a complete basis). Majumdar and 
Ghosh later performed a similar calculation for the one-dimensional spin chain with nearest- and 
next-nearest-neighbour interactions.~\cite{Majumdar69} 

The valence bond basis has a special 
connection to the paramagnetic states of interacting antiferromagnets. Fazekas and Anderson 
introduced the resonating valence bond (RVB) picture to describe a possible spin liquid in 
frustrated magnets.~\cite{Fazekas74} Anderson later proposed that a doped RVB state may 
describe the cuprate superconductors.~\cite{Anderson87a,Anderson87b} This suggestion spurred 
wide interest in the VB description of antiferromagnets and possible exotic ground states 
that are naturally described in terms of VB states. In particular, VB states have proved 
useful as variational states, where the variational freedom lies in the distribution of 
valence bond lengths~\cite{Liang88,Kohmoto88,Lou06} or in a set of projected-BCS 
coefficients.~\cite{Poilblanc89,Gros88,Capriotti01} 

Since valence bond states have total 
spin invariance built in, they are also a natural choice for exact diagonalization within 
the low-energy $S=0$ matrix block. Such calculations have been used to study RVB states
with a restriction of the VB basis to include only a subset of states with short
dimers,~\cite{Lin90,Ramsesha90} which should be a good approximation for systems with
only short-range spin correlations. A generalization of this procedure---in which the valence 
bonds are antisymmetrized products of individual atomic eigenstates---has been widely adopted 
by computational chemists for use in molecular quantum mechanics.~\cite{Balint69,Soos90}

Variational calculations suggest that the collinear N\'{e}el ground state of the 
$d$-dimensional ($d>1$) Heisenberg antiferromagnet is well-described by a superposition 
of VB configurations whose distribution of bond lengths exhibits $1/r^p$ powerlaw 
behaviour, with $p < 5$ in two dimensions.\cite{Liang88} In two dimensions, it is also
believed that there are magnetically disordered states with energy very close to that 
of the ordered ground state.~\cite{Liang88} Thus, competing interactions may favour RVB 
states with short-ranged bonds and no long-range magnetic order.~\cite{Figueirido89,Bose91} 
They may also favour states with valence bond solid (VBS) order,~\cite{Read89,Read90}
in which the translational symmetry is broken but the spin-rotational symmetry remains intact.
To date, the only firm confirmation of an RVB state is in simplified quantum dimer 
models,~\cite{Kivelson87} in which only the dimer degrees of freedom are retained and the 
spin degrees of freedom associated with the dimers are neglected. 

In most cases, possible exotic spin states~\cite{Sondhi97,Sachdev03,Vishwanath04,Senthil04} 
simply cannot be studied in detail with unbiased methods such as quantum Monte Carlo 
(QMC), because of the infamous negative-sign problem. QMC simulations of quantum spin 
systems have traditionally been carried out in the standard $S^z$ basis. Although the 
feasibility of a Monte Carlo projection for improving a variational VB state was demonstrated 
more than fifteen years ago,\cite{Liang90} the use of the VB basis in QMC studies has been 
very limited so far.\cite{Santoro99} As it turns out, neither the overcompleteness nor the 
nonorthogonality of 
the valence bond basis is an impediment to carrying out QMC simulations in principle. In 
fact, even without a good variational state as a trial state, the ground state can be 
completely projected out starting with, \emph{e.g.}, an arbitrarily chosen basis state~\cite{Sandvik05}
using importance sampling and a simple local updating scheme. This method delivers 
performance comparable to the current state of the art for $S^z$-basis QMC calculations
(\emph{i.e.}, stochastic series expansion~\cite{Syluasen02} and world line~\cite{Evertz03} 
methods with loop-cluster updates). It was also noted in Ref.~\onlinecite{Sandvik05} 
that the VB projector method expands  the class of models that are sign-problem-free. 
A host of isotropic SU(2) invariant models with multi-spin interactions (involving 
four, six, \emph{etc.}~spins) can be studied, which in spite of not being frustrated in the standard 
sense (an odd number of antiferromagnetic interactions around a closed loop on the lattice) 
do give rise to a sign problem in standard methods. Thus, VB simulations open new 
opportunities to study ground state phases and transitions in quantum spin systems.

In addition to these new opportunities for QMC studies in the VB basis, we also 
believe that there is more to do variationally. There was not much follow-up on
the pioneering calculations by Liang, Doucot, and Anderson,\cite{Liang88} and with 
the increase in computer performance since that time, it is now feasible to consider more
complex wave-function optimizations.~\cite{Lou06} Although frustrated systems also 
cause sign problems in variational calculations, it may still be possible to extract useful 
information from them. It may also prove fruitful to study variational VB wave functions 
for nonfrustrated multi-spin interactions.

In both QMC and variational calculations, one would like to study a wide range of physical 
observables to characterize the ground state. Overlaps and matrix elements between two 
VB states $\lvert v\rangle$ and $\lvert v'\rangle$ can be related to the structure of the closed loops 
that are formed when their corresponding dimer configurations are superimposed.
There are two standard results:
(i) For a system of $2N$ spins,  $\langle v | v' \rangle = \pm 2^{N_{\circlearrowleft}-N}$
where $N_{\circlearrowleft} \le N$ is the number of loops. 
$\langle v | v' \rangle$ is unity when the states $\lvert v \rangle$ 
and $\lvert v' \rangle$ are identical and $N_{\circlearrowleft}$ is a maximum; its magnitude 
is halved each time a bond mismatch reduces the loop count by 
one.~\cite{Rokhsar88,Sutherland88}
(ii) $\langle v | \mathbf{S}_i\cdot\mathbf{S}_j | v' \rangle =
\pm \frac{3}{4} \langle v | v' \rangle$ if $i$ and $j$ belong to the same loop
and vanishes otherwise.~\cite{Liang88}
(In each case, the overall sign depends on the convention for assigning directions
to the bonds.)
We are not aware of expressions for more complicated matrix elements, \emph{e.g.}, 
those involving more than two spin operators or individual components of the spins.
Nor are we aware of expressions for quantities that are associated with triplet excitations
or for quantities, such as the spin stiffness, that cannot usually be written in terms of an 
equal-time correlation function. 

The goal of this paper is to provide a formal framework for organizing calculations in the 
valence bond basis and to present several new formulas 
relating the loop structure of overlapping valence bond states to a wider range of physical 
properties of the system. We do this in a formal way, using bond operators equivalent 
to those first derived by Sachdev and Bhatt,~\cite{Sachdev90} which in their usual context 
are associated with a fixed dimer configuration. Here, we generalize the creation and 
annihilation operators that describe the states of the two-spin system to the many-spin case, 
allowing the operators to act between arbitrary pairs of spins. With one additional anticommutator 
rule, we find that we can use these operators to organize calculations in the overcomplete 
valence bond basis. As a result, we are then able to address higher-order spin correlations, 
such as $(\mathbf{S}_i\cdot\mathbf{S}_j)(\mathbf{S}_k\cdot\mathbf{S}_l)$, and demonstrate that 
their matrix elements are topological in nature, depending not only on the loop membership of 
the various site labels (as is the case for $\mathbf{S}_i\cdot\mathbf{S}_j$) but also on the 
overall connectivity of the loops with respect to the $(i,j)$ and $(k,l)$ vertices. We introduce 
a generating function whose derivatives produce a related class of cumulant function. A 
diagrammatic expansion elucidates the structure of these functions and greatly simplifies 
their calculation. We extend the valence bond basis to include triplet as well as singlet 
bonds in an effort to access the full Hilbert space. This allows us to compute 
various spin correlation functions component-wise, including two-spin operators of the form 
$S_i^xS^x_j$ and four-spin operators such as $S_i^xS^x_jS^x_kS^x_l$ and $S_i^xS^x_jS^y_kS^y_l$. 
We also derive an expression for the singlet-triplet gap
and the spin stiffness. These results should be useful in the 
context of QMC, variational calculations, and in exact diagonalization.  Our formalism 
may also find use in approximate analytical calculations.

The organization of the paper is as follows. In Sec.~\ref{SEC:Formalism}, we introduce 
the valence bond operator formalism that is used throughout.
The three subsequent sections review how to construct 
$S=0$ valence bond states (Sec.~\ref{SEC:Basis}), how to describe their
evolution under action by a hamiltonian (Sec.~\ref{SEC:Evolution}),
and how to compute their overlaps (Sec.~\ref{SEC:Overlaps}).
In Sec.~\ref{SEC:Correlation}, we begin to derive formulas for the isotropic
spin correlation functions in a somewhat naive way. We reproduce these results in 
Sec.~\ref{SEC:Cumulant} using a more sophisticated diagrammatic approach and then carry 
the calculation to higher order. In Sec.~\ref{SEC:Extended}, we develop rules for
valence bond states with triplet excitations; these are used in Sec.~\ref{SEC:Components} 
to compute correlation functions of  components of the staggered magnetization.
In Sec.~\ref{SEC:Neel}, we discuss the N\'{e}el state and how it can be employed
as a reference state to compute the singlet-triplet gap (Sec.~\ref{SEC:Singlet-triplet})
and the spin stiffness (Sec.~\ref{SEC:Stiffness}).

\section{\label{SEC:Formalism} Valence bond operator formalism}

Two SU(2) spins labelled $i$ and $j$ have a total spin $\mathbf{S} = \mathbf{S}_i+\mathbf{S}_j$
satisfying 
$\tfrac{1}{2}\mathbf{S}^2 = \tfrac{3}{4} + \mathbf{S}_i\cdot\mathbf{S}_j$.
Thus, the usual Heisenberg interaction can be written as
\begin{equation} \label{EQ:Hij}
\hat{H}_{ij} = \mathbf{S}_i\cdot\mathbf{S}_j - \frac{1}{4} = \frac{1}{2}\mathbf{S}^2 - 1.
\end{equation}
The eigenstates of Eq.~\eqref{EQ:Hij} are states of well-defined total spin,
having eigenvalues $\tfrac{1}{2}S(S+1)-1$. We
enumerate them as follows: one ($S=0$)
singlet $\lvert 0 \rangle$ and three ($S=1$) triplets $\lvert 1 \rangle$, $\lvert 2 \rangle$, $\lvert 3 \rangle$
with energy $E^{\mu} = -\delta^{\mu0}$.

For bookkeeping purposes, it is helpful to introduce a boson representation for the spins, 
\begin{equation}
\mathbf{S}_i = \frac{1}{2}\sum_{ss'}b^{\dagger}_{is} \bm{\sigma}_{ss'} b_{is'}
\ \ \text{with} \ \ \sum_{s} b^{\dagger}_{is}b_{is} = 1
\end{equation}
(note the single-occupancy constraint),
and to define a valence bond operator
\begin{equation} \label{EQ:chi-def}
\chi^{\mu\dagger}_{ij} = \frac{1}{\sqrt{2}}\sum_{ss'}\tau^{\mu}_{ss'}
b_{is}^{\dagger} b_{js'}^{\dagger}
\end{equation}
that creates eigenstates of $\hat{H}_{ij}$ out of the bosonic vaccuum (\emph{i.e.},
$\lvert \mu \rangle = \chi_{ij}^{\mu\dagger} \lvert \text{vac} \rangle$
and $\chi_{ij}^{\mu}\lvert \text{vac} \rangle = 0$).
The eigenvalue equation $\hat{H}_{ij}\lvert \mu \rangle = E^{\mu}\lvert \mu \rangle$,
now written as
\begin{equation} \label{EQ:eigenequation}
\hat{H}_{ij}\chi^{\mu\dagger}_{ij} \lvert \text{vac} \rangle = 
- \delta^{\mu0} \chi^{0\dagger}_{ij}\lvert \text{vac} \rangle,
\end{equation}
determines the unknown coefficients $\tau^\mu$.
One possible solution to Eq.~\eqref{EQ:eigenequation} is
\begin{equation} \label{EQ:tau}
\tau^\mu = (\tau^0,{\bm \tau}) = (i\sigma^2, i\sigma^3, \mathbb{1}, -i\sigma^1),
\end{equation}
where $\sigma^{\mu} = (\mathbb{1},{\bm \sigma})$ denotes the four-vector
consisting of the unit matrix and the three Pauli matrices; the resulting states are
\begin{equation}
\lvert \mu \rangle = \chi^{\mu\dagger}_{ij}\lvert\text{vac}\rangle = \begin{cases}
\frac{1}{\sqrt{2}}(\lvert \uparrow_i\downarrow_j \rangle - \lvert \downarrow_i\uparrow_j \rangle )
& \text{if $\mu = 0$}\\
\frac{i}{\sqrt{2}}(\lvert \uparrow_i\uparrow_j \rangle - \lvert \downarrow_i\downarrow_j \rangle)
& \text{if $\mu = 1$}\\
\frac{1}{\sqrt{2}}(\lvert \uparrow_i\uparrow_j \rangle + \lvert \downarrow_i\downarrow_j \rangle)
& \text{if $\mu = 2$}\\
\frac{-i}{\sqrt{2}}(\lvert \uparrow_i\downarrow_j \rangle + \lvert \downarrow_i\uparrow_j \rangle)
& \text{if $\mu = 3$.}
\end{cases}
\end{equation}

Other linear combinations of the triplet states are equally valid, but this choice has the advantage 
that the labels correspond to real physical directions in the standard basis of $\mathbb{R}^3$. 
This is true in the sense that $S^a_i \lvert 0 \rangle \sim \lvert a \rangle$ for $a=1,2,3$.
The key is that Eq.~\eqref{EQ:tau} obeys ${\bm \sigma}\tau^0 = i{\bm \tau}$, 
which implies that
$\mathbf{S}_i\chi_{ij}^{0\dagger}\lvert \text{vac} \rangle
= (i/2) {\bm \chi}_{ij}^{\dagger}\lvert \text{vac} \rangle$.

We emphasize that the $\chi_{ij}^{\mu}$ operators completely describe the 
$\text{SU(2)}\otimes\text{SU(2)} \simeq \text{SO(4)}$ degrees of freedom
of the two-spin system. These operators
obey the completeness relation
\begin{equation} \label{EQ:chi-completeness}
\sum_{\mu} \chi_{ij}^{\mu\dagger}\chi_{ij}^{\mu} =\chi_{ij}^{0\dagger}\chi_{ij}^{0} + \bm{\chi}_{ij}^{\dagger}\cdot\bm{\chi}_{ij} = 1,
\end{equation}
which follows from 
$\sum_{\mu}\tau_{ss'}^\mu\tau^{\mu\dagger}_{r'r}
 = 2\delta_{sr}\delta_{s'r'}$ and the restriction to one boson per site.
The anticommutation relation
\begin{equation} \label{EQ:chi2-commutator}
[\chi^{\mu}_{ij},\chi^{\nu\dagger}_{ij}]\lvert \text{vac} \rangle = \delta^{\mu\nu}\lvert \text{vac} \rangle
\end{equation}
is inherited from the properties 
$[b_{is},b_{js'}^\dagger] = \delta_{ij}\delta_{ss'}$ and $b_{is}\lvert \text{vac} \rangle = 0$.

Any valid (\emph{i.e.}, $\sum_s b_{is}^\dagger b_{is}=$1 preserving) operation on one or both of the spins has an SO(4) representation.
By construction, the operator equivalence
\begin{equation} \label{EQ:SdotS-chi}
\mathbf{S}_i \cdot \mathbf{S}_j = -\frac{3}{4} \chi_{ij}^{0\dagger}\chi_{ij}^{0}
+ \frac{1}{4} \bm{\chi}_{ij}^{\dagger}\cdot\bm{\chi}_{ij}
\end{equation}
holds, and elimination of ${\bm \chi}$ via Eq.~\eqref{EQ:chi-completeness} yields
\begin{equation} \label{EQ:SdotS-quarter-chi}
-\hat{H}_{ij} = \frac{1}{4} - \mathbf{S}_i \cdot \mathbf{S}_j = \chi_{ij}^{0\dagger}\chi_{ij}^{0}.
\end{equation}
For an arbitrary bilinear operator $\hat{O} = \sum_{\mu\nu} O^{\mu\nu}\chi^{\mu\dagger}_{ij}
\chi^{\nu}_{ij}$,
two applications of Eq.~\eqref{EQ:chi2-commutator}
will coax out the coefficient matrix,
$O^{\mu\nu} = \langle\text{vac} \lvert [ \chi^\mu, [\hat{O},\chi^{\nu\dagger}]]
\lvert \text{vac} \rangle$.

In the case of the spin operators themselves, one finds
$(S^a_i)^{\mu\nu} = \frac{1}{2}\tr(\tau^{\mu\dagger}\sigma^a\tau^\nu)$
and
$(S^a_j)^{\mu\nu} = \frac{1}{2}\tr(\tau^{\nu}\sigma^{a*}\tau^{\mu\dagger})$.
Evaluation of the traces leads to
\begin{align} \label{EQ:SiminusSj}
S^a_i - S^a_j &= i\bigl( \chi^{a\dagger}_{ij} \chi^0_{ij} - 
\chi^{0\dagger}_{ij}\chi^a_{ij}\bigr),\\ \label{EQ:SiplusSj}
S^a_i + S^a_j &= i\epsilon^{abc}\chi^{b\dagger}_{ij} \chi^c_{ij}.
\end{align}
Equations~\eqref{EQ:SiminusSj} and \eqref{EQ:SiplusSj} turn out to be 
very useful, but we should not regard them as the fundamental operator equivalence 
rules. They seem to suggest that two-spin operations are always quartic in $\chi^{\mu}_{ij}$,
which is clearly not true [\emph{cf.}\ Eq.~\eqref{EQ:SdotS-chi}].

There is an alternative to computing the coefficient matrix directly.
For operators that transform in a known way, we can induce the same
transformation in $\chi^{\mu}_{ij}$ and equate the corresponding terms.
Consider a spin rotation $S^a \to R^{ab}(\theta\mathbf{n})S^b$
of $\theta$ radians about the axis $\mathbf{n}$ ($\lvert \mathbf{n} \rvert = 1$).
In the boson language, this rotation is equivalent to 
the unitary transformation $b \to Ub$ with
$U(\theta\mathbf{n}) = e^{(i\theta/2)\mathbf{n}\cdot\sigma} = \mathbb{1}\cos(\theta/2) + i\mathbf{n}
\cdot {\bm \sigma} \sin(\theta/2)$.
Now suppose that we rotate $\mathbf{S}_i$ and $\mathbf{S}_j$ by two different angles
about the same axis: putting $b_i \to U(\theta_i\mathbf{n})b_i$ 
and $b_j \to U(\theta_j\mathbf{n})b_j$
into Eq.~\eqref{EQ:chi-def}, 
we find that  the rotation 
is equivalent to the valence bond operator transformation
\begin{equation} \label{EQ:rottransform}
\chi^0_{ij} \to \chi^0_{ij}\cos(\theta_{ij}/2) + \mathbf{n}\cdot{\bm \chi}_{ij}\sin(\theta_{ij}/2),
\end{equation}
where $\theta_{ij} = \theta_i - \theta_j$ is the relative rotation angle.

If $\mathbf{n}$ is directed along the 3 axis, the rotation matrix
$R^{ab} = \frac{1}{2}\tr U^\dagger \sigma^a U \sigma^b$
has the form
\begin{equation}
R(\theta\mathbf{e}^3) = \begin{pmatrix} \cos\theta & \sin\theta & 0 \\
-\sin\theta & \cos\theta & 0 \\
0 & 0 & 0
\end{pmatrix}.
\end{equation}
Described in terms of raising and lowering operators, the transformation amounts to
\begin{equation}
\begin{split}
S^{+} & \to e^{-i\theta}S^{+}\\
S^{-} & \to e^{i\theta}S^{-}\\
S^3 & \to S^3,
\end{split}
\end{equation}
and in particular,
\begin{equation}
S^+_iS^-_j + S^-_iS^+_j \to e^{-i\theta_{ij}} S_i^+S_j^-
+ e^{i\theta_{ij}} S_i^+S_j^-.
\end{equation}
Consequently, the isotropic Heisenberg interaction, expanded in powers of $\theta_{ij}$, behaves as
\begin{multline} \label{EQ:SidotSjrotate}
\mathbf{S}_i\cdot\mathbf{S}_j \to \mathbf{S}_i\cdot\mathbf{S}_j -\frac{i}{2}\theta_{ij} \Bigl( S_i^+S_j^-
- S_i^+S_j^-\Bigr)\\ - \frac{1}{4}\theta_{ij}^2\Bigl( S_i^+S_j^-
+ S_i^-S_j^+\Bigr).
\end{multline}

According to Eq.~\eqref{EQ:rottransform}, 
the same transformation applied to the valence bond operators gives
\begin{multline} \label{EQ:chidaggerchirotate}
\chi^{0\dagger}_{ij}\chi^{0}_{ij} \to \chi^{0\dagger}_{ij}\chi^{0}_{ij} +
\frac{\theta_{ij}}{2}\bigl( \chi^{0\dagger}_{ij}\chi^{3}_{ij}
+ \chi^{3\dagger}_{ij}\chi^{0}_{ij} \bigr)\\
+ \frac{\theta_{ij}^2}{4}\bigl( -\chi^{0\dagger}_{ij}\chi^{0}_{ij} 
+\chi^{3\dagger}_{ij}\chi^{3}_{ij} \bigr).
\end{multline}
Comparing Eqs.~\eqref{EQ:SidotSjrotate} and \eqref{EQ:chidaggerchirotate}
allows us to make the identification
\begin{equation}
\begin{split}
i\bigl( S_i^+S_j^- - S_i^+S_j^-\bigr) &= \chi^{0\dagger}_{ij}\chi^{3}_{ij}
+ \chi^{3\dagger}_{ij}\chi^{0}_{ij}, \\
S_i^+S_j^- + S_i^+S_j^- &= -\chi^{0\dagger}_{ij}\chi^{0}_{ij}
+ \chi^{3\dagger}_{ij}\chi^{3}_{ij}.
\end{split}
\end{equation}

\begin{table}
 \caption{\label{TAB:Opequiv} Operator equivalence rules}
\begin{ruledtabular}
\begin{tabular}{@{\qquad\qquad}c@{\ \ }|c@{\qquad\qquad}}
spin basis & valence bond basis
 \tabularnewline \hline
$\mathbf{S}_i\cdot\mathbf{S}_j$ & $-\frac{3}{4}\chi_{ij}^{0\dagger}\chi_{ij}^0
+ \frac{1}{4}{\bm \chi}_{ij}^\dagger\cdot{\bm \chi}_{ij}$
 \tabularnewline  
$\mathbf{S}_i\cdot\mathbf{S}_j-\frac{1}{4}$ & $-\chi_{ij}^{0\dagger}\chi_{ij}^0$
 \tabularnewline  
$\mathbf{S}_i - \mathbf{S}_j$ & $i( \chi_{ij}^{0\dagger}{\bm \chi}_{ij}
-{\bm \chi}_{ij}^{\dagger}\chi_{ij}^0)$
 \tabularnewline  
$\mathbf{S}_i + \mathbf{S}_j$ & $i {\bm \chi}_{ij}^{\dagger}\times {\bm \chi}_{ij}$
 \tabularnewline  
 $S^+_iS^-_j + S^-_iS^+_j$ & $-\chi_{ij}^{0\dagger}\chi_{ij}^0 
 + \chi_{ij}^{3\dagger}\chi_{ij}^3$
 \tabularnewline  
 $i(S^+_iS^-_j - S^-_iS^+_j)$ & $\chi_{ij}^{0\dagger}\chi_{ij}^3 
 + \chi_{ij}^{3\dagger}\chi_{ij}^0$
\end{tabular}
\end{ruledtabular}
\end{table}

We have so far confined our discussion to a system of two spins.
Nonetheless, everything derived up to this point applies equally well to \emph{any} two spins of
a many-spin system. In that sense, the results summarized in Table~\ref{TAB:Opequiv} 
are valid generally: we simply assert that there are operators $\chi^{\mu}_{ij}$ associated with
\emph{every} pair of sites in the lattice. The only remaining step is to determine what 
the appropriate anticommutator algebra is for operators with one site index in 
common. A straightforward calculation shows that
\begin{equation} \label{EQ:chi3-commutator}
[\chi^{\rho}_{ij},\chi^{\mu\dagger}_{kj}\chi^{\nu\dagger}_{il}]\lvert \text{vac} \rangle = \frac{1}{2}\,T^{\lambda\mu\rho\nu}\,\chi^{\lambda\dagger}_{kl}
\lvert \text{vac} \rangle,
\end{equation}
where 
$T^{\lambda\mu\rho\nu} = \frac{1}{2}\tr \tau^{\lambda\dagger} \tau^{\mu} \tau^{\rho\dagger} \tau^\nu$
and summation over the repeated index $\lambda$ is implied.

\section{\label{SEC:Basis} $S=0$ valence bond basis}

For a system of many SU(2) spins, the structure of the Hilbert space
can be determined from the product rule
$\tfrac{1}{2} \otimes S = (S-\tfrac{1}{2})\oplus(S+\tfrac{1}{2})$, where $S$
denotes the $(2S+1)$-degenerate spin-$S$ state.
Thus,
\begin{equation}
\begin{split}
\tfrac{1}{2}\otimes\tfrac{1}{2} &= 0\oplus 1 \\
\tfrac{1}{2}\otimes\tfrac{1}{2}\otimes\tfrac{1}{2} & = 
\tfrac{1}{2}\oplus\tfrac{1}{2}\oplus \tfrac{3}{2}\\
\tfrac{1}{2}\otimes\tfrac{1}{2}\otimes\tfrac{1}{2}\otimes\tfrac{1}{2} &= 0 \oplus 0 \oplus 1 \oplus 1 \oplus 1 \oplus 2\\
&\,\,\,\vdots \\
\underbrace{\tfrac{1}{2}\otimes\tfrac{1}{2}\otimes\tfrac{1}{2}\otimes\cdots\otimes\tfrac{1}{2}}_{2N\ \text{times}} &=
\prod_{S=0}^N \underbrace{S\oplus\cdots\oplus S}_{C^{(N)}_S\ \text{times}}
\end{split}
\end{equation}
The number, $C^{(N)}_{S}$, of $S$-blocks that appears in the Hilbert space of $2N$ 
spins is given in Table~\ref{EQ:block-count}. The number of states in the
singlet sector, $C^{(N)}_{0} = \frac{1}{N+1}{2N \choose N} = \frac{(2N)!}{N!(N+1)!}$,
represents a small fraction of the total number of states,
$\sum_{S=0}^{N} (2S+1) C^{(N)}_{S} = 2^{2N}$.

\begin{table}[h]
 \caption{ \label{EQ:block-count} Values of the coefficient $C^{(N)}_S$,
 representing the number of spin-$S$ blocks in
 the Hilbert space of $N$ pairs of SU(2) spins.}
\begin{ruledtabular}
\begin{tabular}{@{\quad}l|cccccccccc}
 & 0 & 1/2 & 1 & 3/2 & 2 & 5/2 & 3 & 7/2
 \tabularnewline \hline
 & & 1 \tabularnewline
1 &  1 & &  1 \tabularnewline
 &   & 2 & & 1  \tabularnewline
2 &  2 & & 3 & & 1 \tabularnewline
 & & 5  & & 4 & & 1\tabularnewline
3 & 5 && 9  && 5 && 1 \tabularnewline
 &&  14 &&  14 && 6 && 1 \tabularnewline
4 & 14 && 28 && 20  && 7 && 1 \tabularnewline
 &&  42 &&  48 && 27 && 8 && 1 \tabularnewline
5 & 42 && 90 && 75  && 35 && 9 \tabularnewline
  && 132 && 165 && 120  && 24 && 10 \tabularnewline
6 & 132 && 297 && 285  && 144 && 34 \tabularnewline  
\end{tabular}
\end{ruledtabular}
\end{table}

To construct a valence bond state in the singlet sector
[as per Eq.~\eqref{EQ:vbstate}], we simply
group the spins into $N$ pairs and, starting from the bosonic vacuum,
act with $\chi^{0\dagger}_{ij}$ for each pair $(i,j)$.
The number of such states is
\begin{equation}
T_N = \frac{1}{N!} {2N \choose 2}{2N-2 \choose 2}\cdots{2 \choose 2} = \frac{(2N)!}{2^N(N!)}.
\end{equation}
This dwarfs the number of independent states in the singlet sector
for large $N$ ($T_N \gg C^{(N)}_0$).

As a consequence of the overcompleteness, linear combinations of valence bond 
states are generally not  unique---an ambiguity related to the fact that linear independence fails 
even at the level of two singlets:
\begin{equation} \label{EQ:linearly-dependent}
\chi^{0\dagger}_{il}\chi^{0\dagger}_{jk}
+ \chi^{0\dagger}_{ij}\chi^{0\dagger}_{kl}
+ \chi^{0\dagger}_{lj}\chi^{0\dagger}_{ik} = 0.
\end{equation}
In the case of two spins ($N=1$), there is a single tiling ($T_1 = 1$) corresponding to the
unique singlet state ($C^{(1)}_0 = 1$). In the case of four spins ($N=2$), the number of 
tilings ($T_2 = 3$) exceeds the number of physical singlet states ($C^{(2)}_0 = 2$) by one.
It is useful to eliminate this superfluous state. For concreteness, 
let us suppose that the lattice of $2N$ spins is described as the union
$A\cup B$ of sites $A = \{ i_1, i_2, \ldots, i_N\}$
and $B = \{ j_1, j_2, \ldots, j_N\}$.
This is simply a labelling trick and does not depend on the
lattice being bipartite. We restrict the basis to include only
those valence bonds that connect A sites to B sites.
This is possible since any unwanted AA and BB bonds can be replaced 
by two AB bonds via Eq.~\eqref{EQ:linearly-dependent}.

\begin{figure}
\includegraphics{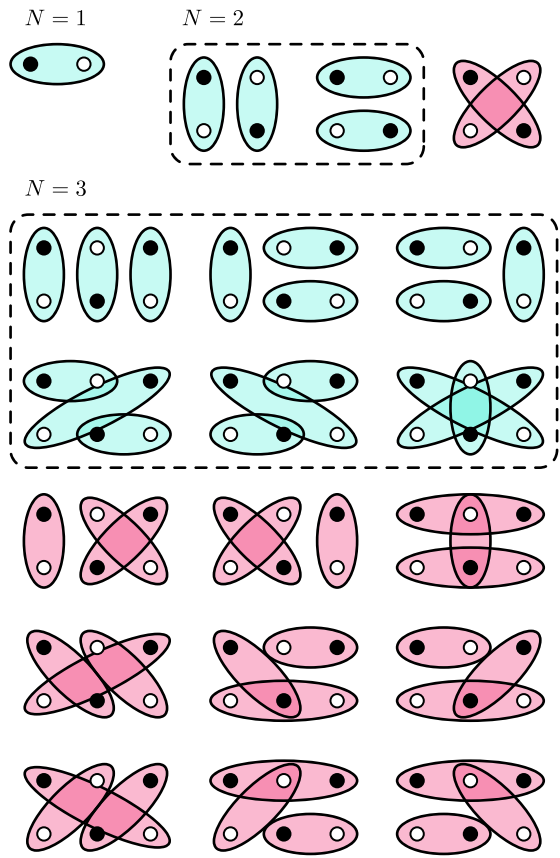}
\caption{\label{FIG:tilings}The set of all possible bond tilings,
$T_N = (2N)!/2^NN!$ in number, is shown for $N=1,2,3$.
The restricted set of AB tilings (blue) numbers $\tilde{T}_N = N!$,
which still exceeds the number of true singlet states.
A and B sites are shown as filled and open circles, respectively.
}
\end{figure}

The AB valence bond basis is intimately related to $\mathcal{S}_N$, the
symmetric group of degree $N$. There is a one-to-one correspondence between the bond
tilings and the permutations that are the elements of $\mathcal{S}_N$.
This correspondence holds because
every bond tiling can be thought of as a rearrangement of the $B$ labels.
That is, for each $P \in \mathcal{S}_N$, there is a state
\begin{equation} \label{EQ:operator-string}
\lvert P \rangle = \hat{P}^\dagger\lvert \text{vac}\rangle = \prod_{n=1}^N\chi^{0\dagger}_{i_n,j_{Pn}}\lvert \text{vac} \rangle.
\end{equation}
It follows that there are $\tilde{T}_N = N!$ such states.
For example, Fig.~\ref{FIG:tilings} shows the $3! = 6$ AB bond configurations
for $N=3$. These states are indexed by the permutations
\begin{align*}
P &=  (1)(2)(3) & P &=  (1)(2\ 3) & P &=  (1\ 2)(3) \\
P &= (1\ 2\ 3) &  P &= (3\ 2\ 1) & P &= (1\ 3)(2),
\end{align*}
written here in cycle notation.

The restriction to the AB basis is useful in that it gives us a
systematic way to fix the phase of each state.
This is an issue because the singlet is a directed bond: $\chi^{0}_{ji} = -\chi^{0}_{ij}$
and ${\bm \chi}_{ji} = {\bm \chi}_{ij}$.
We adopt the convention that the canonical form of the operator 
$\chi^\mu_{ij}$ has $i\in A$
and $j \in B$, as in Eq.~\eqref{EQ:operator-string}.
As we shall see in Sect.~\ref{SEC:Overlaps}, all overlaps of AB valence bond states are
positive definite, which is not true in the unrestricted basis.

One disadvantage to the AB basis is that the identity operator
in the $S=0$ subspace has a complicated form.
In the unrestricted basis, it is diagonal,
\begin{equation} \label{EQ:identityop}
\hat{1}_0 = \frac{2^N}{(N+1)!}\sum_{v} \lvert v \rangle \langle v \rvert,
\end{equation}
with a  normalization $T_N/C^{(N)}_0$\!, but the process of eliminating all AA and BB 
bonds from Eq.~\eqref{EQ:identityop} introduces offdiagonal terms.
For $N=2$ and $N=3$, the identity operators are
\begin{equation}
\hat{1}_0 = \frac{2}{3} \sum_{P,Q}\lvert P \rangle
\begin{pmatrix} \phantom{+}2 & -1 \\ -1 & \phantom{+}2 \end{pmatrix}_{\!\!P,Q}
\!\!\!\!\!\!\langle Q \rvert
\end{equation}
and
\begin{equation}
\hat{1}_0 = \frac{1}{3}\sum_{P,Q}\lvert P \rangle
\begin{pmatrix} \phantom{+}4 & -1 & -1 & \phantom{+}0 & \phantom{+}0 & -1 \\
-1 & \phantom{+}4 & \phantom{+}0 & -1 & -1 & \phantom{+}0 \\ 
-1 & \phantom{+}0 & \phantom{+}4 & -1 & -1 & \phantom{+}0 \\ 
\phantom{+}0 & -1 & -1 & \phantom{+}4 & \phantom{+}0 & -1 \\ 
\phantom{+}0 & -1 & -1 & \phantom{+}0 & \phantom{+}4 & -1 \\
-1 & \phantom{+}0 & \phantom{+}0 & -1 & -1 & \phantom{+}4
\end{pmatrix}_{\!\!P,Q}
\!\!\!\!\!\!\langle Q \rvert. 
\end{equation}
In these examples, the rows and columns of the coefficient matrices are arranged to match
the ordering of the permutations in Fig.~\ref{FIG:tilings}. The entries are 
$(2N)!/2^{N-1}(N!)^2-1$ along the main diagonal and $0$ or $-1$ elsewhere, which is
a consequence of there being at most one AA/BB set to unravel. For $N>3$,
the situation is more complicated, and we do not know of a simple general 
expression for $\hat{1}_0$. In most practical applications, however, such an 
expression is not needed.

It is possible to construct basis sets that are still more restrictive
(\emph{i.e.}, having fewer states than the AB basis).
There is a hierarchy of linear dependence relations 
[similar to Eq.~\eqref{EQ:linearly-dependent}] for groups
of three, four, and higher numbers of singlet bonds, which can
be used to further eliminate unwanted states.
When carried out to its fullest extent, this elimination procedure
leaves a set of valence bond configurations equal in number to the 
actual number of $S=0$ states. Explicit construction of the basis
in this limit can be accomplished by arranging the lattice sites 
$i_1,j_1,i_2,j_2,\cdots,i_N,j_N$ on a ring and keeping
only the AB tilings that produce no bond crossings.
Note that even this minimal set contains long bonds
connecting sites that are macroscopically separated, which
suggests that bonds on all length scales are required for completeness.

\section{\label{SEC:Evolution} Evolution of valence bond states}

Now suppose that we have a hamiltonian of the form
\begin{equation} \label{EQ:modelH}
\hat{H} = \sum_{ij} J_{ij}\hat{H}_{ij} 
- \sum_{ijkl} K_{ijkl} \hat{H}_{ij} \hat{H}_{kl} + \cdots
\end{equation}
with second-, forth-, and potentially higher-order spin interations.~\cite{Raman05}
Here, $\hat{H}_{ij}$ is defined as in Eq.~\eqref{EQ:SdotS-quarter-chi}.
In order to understand how an arbitrary state evolves under $\hat{H}$,
we need to know how a given valence bond state evolves under 
repeated applications of $-\hat{H}_{ij}$.

Depending on the circumstances, $-\hat{H}_{ij}$
is either a diagonal operation that  leaves the overall configuration unchanged
or an off-diagonal operation that maps two bonds to their complementary tiling:
\begin{align} \label{EQ:updaterule1}
\Bigl(\frac{1}{4}-\mathbf{S}_i\cdot\mathbf{S}_j\Bigr)[i,j] &= [i,j],\\ \label{EQ:updaterule2}
\Bigl(\frac{1}{4}-\mathbf{S}_i\cdot\mathbf{S}_j\Bigr)[l,i][j,k] &= [i,j][k,l].
\end{align}
This result, which is true in the unrestricted basis, has long been known.~\cite{Hulthen38}
It will be instructive to see how it emerges within our formalism
and how it is modified to accommodate the restriction to AB bonds.

Let us start with the state $\lvert P \rangle = \hat{P}^\dagger\lvert \text{vac}\rangle$,
where $\hat{P}^\dagger$ is the operator string defined in Eq.~\eqref{EQ:operator-string}.
Then,
\begin{equation} \label{EQ:updatePhat}
-\hat{H}_{ij}\lvert P \rangle = 
\chi_{ij}^{0\dagger}\chi_{ij}^{0}\hat{P}^\dagger\lvert \text{vac}\rangle
= \chi_{ij}^{0\dagger}[\chi_{ij}^{0}, \hat{P}^{\dagger}] \lvert\text{vac}\rangle.
\end{equation}
There are two possibilities 
to consider in evaluating the anticommutator: 
(i) If there is already a bond connecting the active sites 
(\emph{i.e.}, $\hat{P}^\dagger = \cdots \chi^{0\dagger}_{ij} \cdots$),
then only two operators play a role in the anticommutator. In this case,
Eq.~\eqref{EQ:chi2-commutator} is the appropriate rule to apply, and 
Eq.~\eqref{EQ:updatePhat} simplifies to
\begin{equation} \label{EQ:nonupdated-op-seq}
\chi_{ij}^{0\dagger}\chi_{ij}^{0}\lvert P \rangle = \lvert P \rangle.
\end{equation}
(ii)
If the active sites are each connected elsewhere 
(\emph{i.e.}, $\hat{P}^\dagger = \cdots \chi^{0\dagger}_{il}\cdots\chi^{0\dagger}_{kj} \cdots$ 
for some $k\in A$, $l \in B$), then
three operators are involved. This necessitates the use
of Eq.~\eqref{EQ:chi3-commutator} and leads to 
\begin{equation} \label{EQ:updated-op-seq}
\chi_{ij}^{0\dagger}\chi_{ij}^{0}\lvert P \rangle = 
\frac{1}{2} \chi_{i_1j_{P1}}^{0\dagger}\cdots
\chi_{ij}^{0\dagger}\cdots\chi^{0\dagger}_{kl} \cdots
 \chi_{i_Nj_{PN}}^{0\dagger}
 \lvert \text{vac} \rangle.
\end{equation}
The new state on the right-hand side is itself a valence bond state
(different from $\lvert P \rangle$), since it consists of $N$ operators and 
none of its $2N$ indices are repeated. If $i\in A$ and $j\in B$,
then this state contains only AB bonds.

A reconfiguration of the bonds manifests itself as change in the permutation
indexing the state. If $i = i_n$ and $j = j_m$ are in opposite sublattices,
then Eqs.~\eqref{EQ:nonupdated-op-seq} and \eqref{EQ:updated-op-seq} 
are summarized by the compact update rule
\begin{equation} \label{EQ:updateinjm}
-\hat{H}_{i_nj_m} \lvert P \rangle = \biggl(\frac{1}{2}\biggr)^{\!1-\delta_{Pn,m}}\lvert (Pn\  m) P \rangle,
\end{equation}
where the term $(Pn\ m)$ inside the ket is a 2-cycle acting on $P$
(the effect of which is to swap the labels $j_{Pn}$ and $j_m$).
Equation~\eqref{EQ:updateinjm} makes clear that the hamiltonian is an identity
operation whenever it acts on a preexisting bond; otherwise it induces a single 
transposition and a multiplicative factor $\frac{1}{2}$. Note that the transposition
itself depends on $P$.

\begin{figure}
\begin{center}
\includegraphics{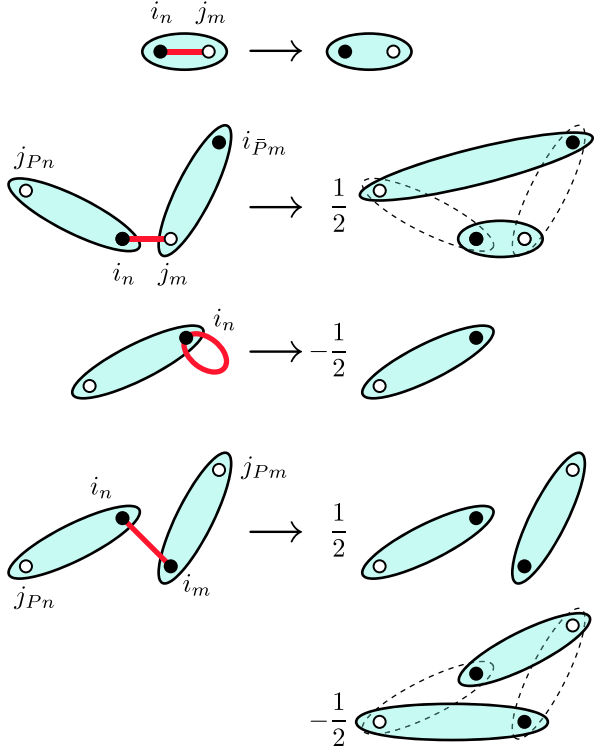}
\end{center}
\caption{ \label{FIG:updates}
AB valence bonds obey a simple set of update rules when acted on with an isotropic
spin interaction term. Here the update rules for
$1/4-\mathbf{S}_i \cdot \mathbf{S}_j$ are summarized.
The red bar denotes the interaction between the spins at sites $i$ and $j$.
The site labels follow the notation in Eqs.~\eqref{EQ:updateinjm}
and \eqref{EQ:updateinim}. Of the terms that involve a reconfiguration
of the bonds (dashed outlines indicate the previous bond locations),
one corresponds to an exchange of $j_{Pn}$ and $j_m$ and the 
other $j_{Pn}$ and $j_{Pm}$.
}
\end{figure}

If, however, $i$ and $j$ are in the same sublattice, then the 
$\chi_{ij}^{0\dagger}\chi^{0\dagger}_{kl}$ operators in
Eq.~\eqref{EQ:updated-op-seq}
have to be eliminated in favour of AB bonds.
This leads to
\begin{equation} \label{EQ:updateinim}
-\hat{H}_{i_ni_m} \lvert P \rangle = \frac{1}{2}\lvert P \rangle - \biggl(\frac{1}{2}\biggr)^{\!1-\delta_{n,m}}
\lvert (Pn\  Pm) P \rangle
\end{equation}
and
\begin{equation} \label{EQ:updatejnjm}
-\hat{H}_{j_nj_m} \lvert P \rangle = \frac{1}{2}\lvert P \rangle - \biggl(\frac{1}{2}\biggr)^{\!1-\delta_{n,m}}
\lvert (n\  m) P \rangle.
\end{equation}
These ``frustrating'' interactions create a linear superposition of states, rather
than just a rearrangement of bonds. They also introduce bond configurations
with negative weight, violating the conditions of the Marshall sign theorem.~\cite{Marshall55}
Figure~\ref{FIG:updates} illustrates the frustrating and nonfrustrating bond flips that can occur.

If there are no frustrating interactions in the hamiltonian then the
ground state can be written as a superposition of valence bond states
$\lvert \psi \rangle = \sum_P c_P \lvert P \rangle$ with all $c_P \ge 0$.
In that special case, Monte Carlo simulation is sign-problem-free. 

\section{\label{SEC:Overlaps} Overlaps and matrix elements of valence bond states}

\begin{figure}
\includegraphics{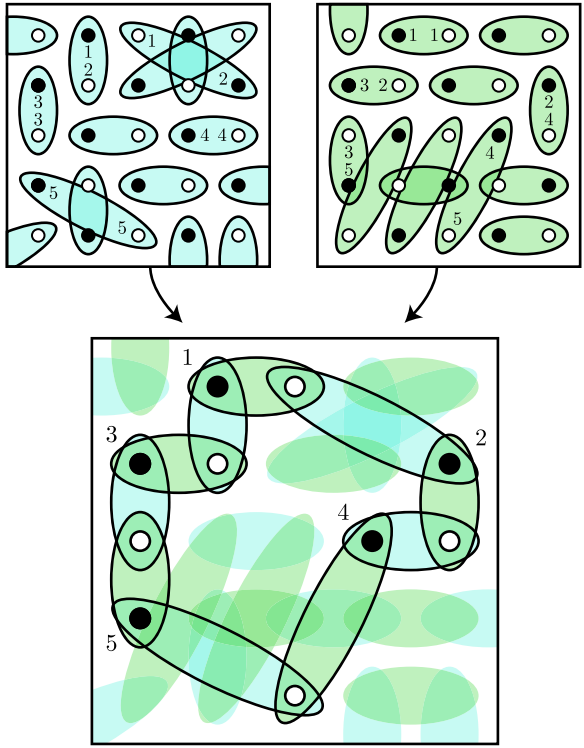}
\caption{ \label{FIG:overlap}
The top-left and top-right panels illustrate valence bond states
$\lvert P \rangle$ and $\lvert Q \rangle$ corresponding to the
permutations
$P = (1\ 2)(3)(4)(5)\cdots$ and
$Q = (1)(2\ 4\ 5\ 3)\cdots$;
the sites $i_1,i_2,\ldots,i_5$ (filled circles) and $j_1,j_2,\ldots,j_5$ (open circles)
are numbered accordingly (in no particular order).
The overlap $\langle Q | P \rangle$ between these states
is literally that: a superposition of the two bond configurations. 
Since the end point of one bond is always the starting point of another,
a set of closed bond paths is formed. These are in one-to-one correspondence
with the disjoint cycles of $\bar{Q}P$.
In this example, $\bar{Q}P = (1\ 2\ 4\ 5\ 3)\cdots$
has a cycle of length 5 and a corresponding bond loop of length 10.
}
\end{figure}

It follows from Eq.~\eqref{EQ:operator-string} that the overlap of any two valence bond states
is equal to the vacuum-state expectation value of a length-$2N$ operator string:
\begin{multline}
\langle Q \vert P \rangle = \langle\text{vac}\rvert \chi^{0}_{i_N,j_{QN}}\cdots
 \chi^{0}_{i_2,j_{Q2}}\chi^{0}_{i_1,j_{Q1}}\\
\times \chi^{0\dagger}_{i_1,j_{P1}}\chi^{0\dagger}_{i_2,j_{P2}}\cdots
\chi^{0\dagger}_{i_N,j_{PN}}\lvert \text{vac} \rangle.
\end{multline}
The creation and annihilation operators can be shuffled past one another
so long as they share no indices in common. The maximal such rearrangement
leaves the string grouped into several linked chains of operators; these
are related to the disjoint cycles of the composite permutation $\bar{Q}P$
(we use the notation $\bar{Q} = Q^{-1}$ to denote the inverse permutation of $Q$)
and
define a set of directed loops (with $\bar{Q}P$ defining the proper order), \emph{e.g.},
\begin{multline}
i_1 \to j_{P1} \to i_{\bar{Q}P1} \to j_{P\bar{Q}P1}\\
 \to i_{\bar{Q}P\bar{Q}P1}
\to \cdots \to i_{(\bar{Q}P)^k1} = i_1.
\end{multline}
Figure~\ref{FIG:overlap} illustrates the construction.
Each of these loops is even in length---constituting a chain with equal numbers of
links from $P$ and $Q$. The cycle decomposition is related to the loop membership
of the site indices by 
\begin{equation}\label{EQ:sameloop}
\begin{split} 
n \overset{\bar{Q}P}{\sim} \bar{P}m &\ \Leftrightarrow\ \text{$i_n$ and $j_m$ in same loop}\\ 
n \overset{\bar{Q}P}{\sim} m &\ \Leftrightarrow\ \text{$i_n$ and $i_m$ in same loop}\\ 
\bar{P}n \overset{\bar{Q}P}{\sim} \bar{P}m &\ \Leftrightarrow\ \text{$j_n$ and $j_m$ in same loop}
\end{split}
\end{equation}
We use the notation $x \overset{\bar{Q}P}{\sim} y$ to indicate that $x$ and $y$ are in the same cycle of $\bar{Q}P$.
All the loops taken together form 
a set of closed paths covering the lattice.
Hence,
\begin{equation} \label{EQ:loop-length-sum}
\sum_{k=1}^{N} 2k\cdot n_{k} = 2N,
\end{equation}
where $n_k$ is the number of loops of length $2k$ (or 
the number of $k$-cycles in $\bar{Q}P$).

The value of the overlap can be computed by way of a simple decimation
scheme.~\cite{Rokhsar88,Sutherland88} Each loop can be eliminated by
repeated application of
Eqs.~\eqref{EQ:chi2-commutator} and \eqref{EQ:chi3-commutator}, specialized
to the case of $\rho = \nu = \mu = 0$:
\begin{equation} \label{EQ:chi02-commutator}
[\chi^{0}_{ij},\chi^{0\dagger}_{ij}]\lvert \text{vac} \rangle = \lvert \text{vac} \rangle
\end{equation}
and
\begin{equation} \label{EQ:chi03-commutator}
[\chi^{0}_{ij},\chi^{0\dagger}_{kj}\chi^{0\dagger}_{il}]\lvert \text{vac} \rangle = \frac{1}{2}\chi^{0\dagger}_{kl}
\lvert \text{vac} \rangle.
\end{equation}
For a loop of length $2k$, $k-1$ applications of Eq.~\eqref{EQ:chi03-commutator}
removes $2k-2$ links and yields $k-1$ factors of 1/2. The final two operators are removed
with a single application of Eq.~\eqref{EQ:chi02-commutator}, leaving 
$\langle \text{vac} | \text{vac} \rangle  = 1$.
Hence,
\begin{equation} \label{EQ:QPoverlap}
\langle Q | P \rangle = \prod_{l=1}^N
\biggl(\frac{1}{2^{k-1}}\biggr)^{n_{k}}
= \frac{2^{\sum_{l=1}^N n_{k}}}{2^{\sum_{l=1}^N k\cdot n_{k}}}
= 2^{N_{\circlearrowleft}-N},
\end{equation}
where we have used Eq.~\eqref{EQ:loop-length-sum}
and defined the total number of loops $N_{\circlearrowleft} = \sum_{k=1}^N n_{k}$.

Similar arguments can be used to compute the matrix elements of
a spin-rotation-invariant operator $\hat{O}$.
In general, such an operator acting on a valence bond state 
produces a linear superposition of states with modified bond configurations:
\begin{equation} \label{EQ:OhatP}
\hat{O}\lvert P \rangle = \sum_{P'} O_{P,P'}\lvert P' \rangle.
\end{equation}
(The situation is more complicated if $\hat{O}$ generates states outside 
the $S=0$ sector; this is discussed further in Sect.~\ref{SEC:Extended}.)
Since the basis is overcomplete, the coefficients $O_{P,P'}$ are not unique and
 $O_{P,P'} \neq \langle P \lvert \hat{O} \rvert P' \rangle$, as would be
the case in an orthonormal basis.
Equations~\eqref{EQ:QPoverlap} and \eqref{EQ:OhatP} imply that 
matrix elements are related to changes in the loop number:
\begin{equation} \label{EQ:QOhatP}
\frac{\langle Q \lvert \hat{O} \rvert P \rangle}{\langle Q | P \rangle}
=  \sum_{P'} O_{P,P'} 2^{N_{\circlearrowleft}'-N_{\circlearrowleft}}.
\end{equation}
Here, $N_{\circlearrowleft}$ is the number of cycles in $\bar{Q}P$ and
$N_{\circlearrowleft}'$ the number in $\bar{Q}P'$.

Evaluation of Eq.~\eqref{EQ:QOhatP} requires some general rules for how the number of cycles changes
as $P \to P'$. Since the permutation $P'\bar{P}$ can always be decomposed into a product of transpositions, it 
suffices to consider the situation where the two states differ by a single transposition.
In that case, $N_{\circlearrowleft}' - N_{\circlearrowleft} = \pm 1$ (see the Appendix~\ref{SEC:Permutation}), since a transposition $(n\ m)$ either merges the two cycles that each contain one of
$\bar{P}n$ and $\bar{P}m$ or splits the single cycle that contains them both.
This can be expressed formally as
\begin{equation} \label{EQ:overlapratio}
\frac{\langle Q \lvert (n\ m) P\rangle}{\langle Q | P \rangle} \biggl(\frac{1}{2}
\biggr)^{1-\delta_{n,m}} 
= \begin{cases}
1 & \text{if $\bar{P}n \overset{\bar{Q}P}{\sim} \bar{P}m$ } \\
\frac{1}{4} & \text{otherwise}.
\end{cases}
\end{equation}
Here, the delta function takes care of the possibility that $n=m$, which corresponds
to no transposition at all.

Applying this result to Eqs.~\eqref{EQ:updateinjm}--\eqref{EQ:updatejnjm}
[with the right-hand side of Eq.~\eqref{EQ:overlapratio} interpreted according 
to \eqref{EQ:sameloop}] gives 
\begin{equation} \label{EQ:chidaggerchi}
\frac{\langle Q \lvert \chi^{0\dagger}_{ij} \chi^{0}_{ij} \rvert P \rangle}
{ \langle Q | P \rangle } = \frac{1}{4}\bigl( 1 - 3\epsilon_{ij}\delta^{\alpha_{i}\alpha_{j}} \bigr),
\end{equation}
where $\alpha_i$ and $\alpha_j$ are unique labels for the loops 
passing through sites $i$ and $j$. We have introduced the notation
\begin{equation}
\epsilon_{ij} = \begin{cases}
-1 & \text{if $i,j$ are in different sublattices}\\
+1 & \text{otherwise}.
\end{cases}
\end{equation}

\begin{figure}
\includegraphics{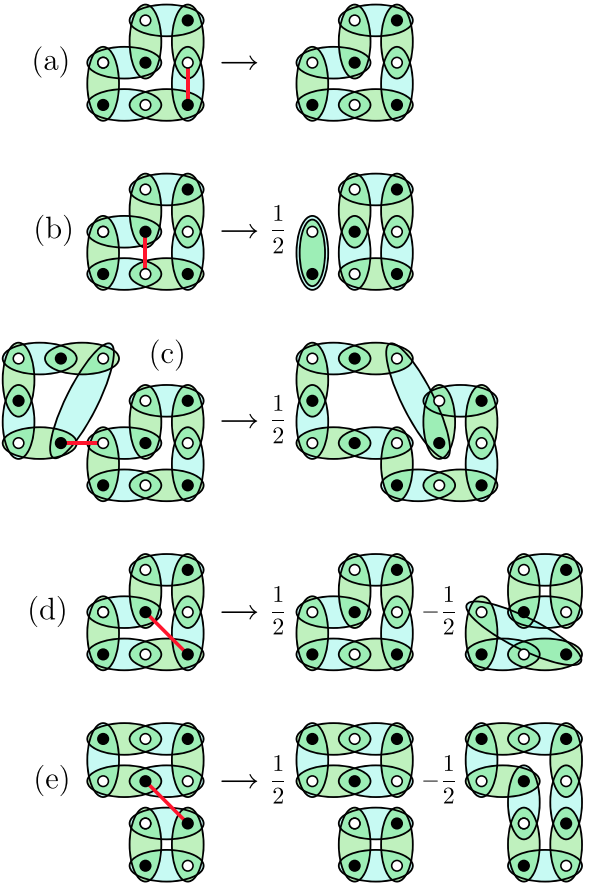}
\caption{ \label{FIG:Cij} 
The operator
$\chi_{ij}^{0\dagger}\chi_{ij}^0 = 1/4-\mathbf{S}_i \cdot \mathbf{S}_j$
causes a rearrangement of bonds, as shown in Fig.~\ref{FIG:updates}.
Thus, its matrix element can be understood in terms of joining
or splitting loops.
}
\end{figure}

To clarify, let us rederive Eq.~\eqref{EQ:chidaggerchi} by considering explicitly
the effect of $\chi^{0\dagger}_{ij}\chi^{0}_{ij}$ acting to the right on $\lvert P \rangle$
and how the overlap of the resulting state with $\langle Q \rvert$ differs from
the loop structure of $\langle Q | P \rangle$.
We distinguish between the AB case, 
in which the sites $i$ and $j$ are in different sublattices, and the AA/BB case, 
in which they are in the same sublattice.
For AB operators, the possibilities are illustrated in Fig.~\ref{FIG:Cij}(a--c).
If the two sites belong to the same loop then either an
offdiagonal operation splits the loop, giving a contribution $(1/2)(2^1) = 1$ [following 
Eq.~\eqref{EQ:QOhatP}]
or a diagonal operation leaves the loop unchanged, also giving $(1)(2^0) = 1$.
If the two sites belong to different loops then the offdiagonal
operation joins the two loops, giving $(1/2)(2^{-1})=1/4$. Combining
these results gives
\begin{equation} \label{EQ:chiijchiij}
\begin{split}
\frac{\langle Q \lvert \chi^{0\dagger}_{ij}\chi^{0}_{ij}
\rvert P \rangle}
{\langle Q | P \rangle}
&= \delta^{\alpha_i\alpha_j} + \frac{1}{4}\Bigl(1-\delta^{\alpha_i\alpha_j}\Bigr)\\
&= \frac{3}{4}\delta^{\alpha_i\alpha_j} + \frac{1}{4},
\end{split}
\end{equation}
For AA/BB operators, the possible bond reconfigurations are illustrated in
Figs.~\ref{FIG:Cij}(d) and \ref{FIG:Cij}(e).
The two sites are either in the same loop, giving
$(1/2)(2^0-2^1) = -1/2$, or in different loops, giving  $(1/2)(2^0-2^{-1}) = 1/4$. The result,
\begin{equation}
\begin{split}
\frac{\langle Q \lvert \chi^{0\dagger}_{ij}\chi^{0}_{ij}
\rvert P \rangle}
{\langle Q | P \rangle}
 &= -\frac{1}{2}\delta^{\alpha_i\alpha_j}+ \frac{1}{4}\Bigl(1-\delta^{\alpha_i\alpha_j}\Bigr)\\
&= -\frac{3}{4}\delta^{\alpha_i\alpha_j} + \frac{1}{4},
\end{split}
\end{equation}
differs from Eq.~\eqref{EQ:chiijchiij} by only a sign. Using $\epsilon_{ij}$
to account for the difference in sign yields Eq.~\eqref{EQ:chidaggerchi}.

The same formal manipulations that lead from Eq.~\eqref{EQ:overlapratio} to 
Eq.~\eqref{EQ:chidaggerchi} can be chained together to evaluate a long 
operator sequence one transposition at a time. For example,
taking $i = i_n \in A$ and $j = j_m \in B$, we get
\begin{equation}
\begin{split}
\frac{\langle Q \lvert \hat{O} \chi^{0\dagger}_{ij}\chi^0_{ij} \rvert P \rangle}
{\langle Q | P \rangle}
&= \frac{\langle Q \lvert \hat{O} \rvert (Pn\ m)P \rangle}
{\langle Q | P \rangle} \biggl(\frac{1}{2}\biggr)^{1-\delta_{Pn,m}} \\
&= \frac{\langle Q \lvert \hat{O} \rvert P' \rangle}
{\langle Q | P' \rangle}
\frac{\langle Q | P' \rangle}
{\langle Q | P \rangle} \biggl(\frac{1}{2}\biggr)^{1-\delta_{Pn,m}} \\
&= \frac{\langle Q \lvert \hat{O} \rvert P' \rangle}
{\langle Q | P' \rangle}
\frac{1}{4}\bigl( 1 + 3\delta^{\alpha_i \alpha_j}\bigr),
\end{split}
\end{equation}
where $P' = (Pn\ m)P$ describes the new bond configuration after one transposition.
For arbitrary site indices, the expression reads
\begin{multline} \label{EQ:Ochiijmatrixelement}
\frac{\langle Q \lvert \hat{O} \chi^{0\dagger}_{ij}\chi^0_{ij} \rvert P \rangle}
{\langle Q | P \rangle}
= \frac{1}{4}(1+\epsilon_{ij})
\frac{\langle Q \lvert \hat{O} \rvert P \rangle}
{\langle Q | P \rangle}\\
- \epsilon_{ij} \frac{\langle Q \lvert \hat{O} \rvert P' \rangle}
{\langle Q | P' \rangle} \frac{1}{4}\bigl( 1 + 3\delta^{\alpha_i \alpha_j}\bigr).
\end{multline}

\section{\label{SEC:Correlation} Spin Correlation Functions}

So long as the ground state of the spin system is a global singlet, its wavefunction can be written
as a linear superposition of valence bond states: $\lvert \psi \rangle = \sum_P c_P\lvert P \rangle$.
Accordingly, operator expectation values have the form
\begin{equation} \label{EQ:psiOpsi}
\langle \hat{O} \rangle = \frac{\langle \psi \lvert \hat{O} \rvert \psi \rangle}{\langle \psi | \psi \rangle}
= \frac{\sum_{P,Q} W(P,Q)
\frac{\langle Q \lvert \hat{O} \rvert P \rangle}{\langle Q | P \rangle}
 }{\sum_{P,Q} W(P,Q) },
\end{equation}
with $P,Q \in \mathcal{S}_N$ and $\hat{O}$ some operator of interest.
The quantity $W(P,Q) = c_Q c_P \langle Q | P \rangle$ may be determined 
variationally~\cite{Lou06} or it may arise as a sampling weight in the context
of a Monte Carlo projection scheme.~\cite{Sandvik05}
The matrix element  $\langle Q | \hat{O} | P \rangle/\langle Q | P \rangle$
is related to the properties of the loops that are formed when the singlet tilings of 
the two states are superimposed. In fact, since its value depends only on the properties
of the loops, the individual valence bond states can be abstracted away entirely,
leaving what is essentially a loop estimator for $\hat{O}$; we denote this function $O_{\mathcal{L}}$.
In this way of thinking, Eq.~\eqref{EQ:psiOpsi} can best be understood as 
$\langle \hat{O} \rangle = \langle O_{\mathcal{L}} \rangle_W$, 
an ensemble average of $O_{\mathcal{L}}$ in the gas of fluctuating 
loops.\cite{Sutherland88b, Kohmoto88b}

\begin{figure}
\includegraphics{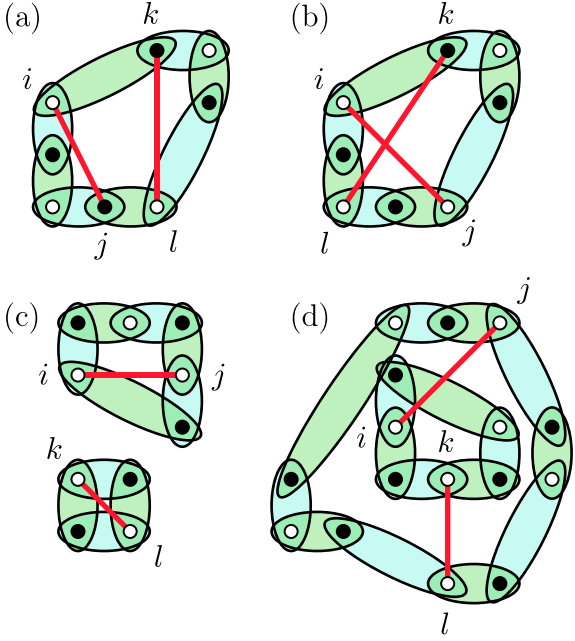}
\caption{ \label{FIG:Cijkl}
The expectation value of
$\epsilon_{ij}\epsilon_{kl}\bigl(\mathbf{S}_i\cdot\mathbf{S}_j\bigr)\bigl(\mathbf{S}_k\cdot\mathbf{S}_l\bigr)$ is nonzero only if $(i,j)$ and $(k,l)$ connect one or two loops.
For one loop, the value is either $9/16$ or $-3/16$ depending on whether
$k$ and $l$ are in the (a) same or (b) different loop segments
connecting $i$ and $j$. For two loops, the value is either $9/16$ or $3/16$
depending on whether the interactions leave the loops (c) disjoint or (d) connected.
}
\end{figure}

In this section, we want to determine the loop estimators for the spin operators $\mathbf{S}_i\cdot\mathbf{S}_j$
and $(\mathbf{S}_i\cdot\mathbf{S}_j)(\mathbf{S}_k\cdot\mathbf{S}_l)$.
All the work to compute the second order result has already been done.
Since $\mathbf{S}_i\cdot\mathbf{S}_j = \frac{1}{4} - \chi^{0\dagger}_{ij}\chi^0_{ij}$,
comparison with Eq.~\eqref{EQ:chidaggerchi} immediately yields
\begin{equation} \label{EQ:SidotSj}
\bigl(\mathbf{S}_i \cdot\mathbf{S}_j\bigr)_{\mathcal{L}}
= \frac{3}{4}\epsilon_{ij} \delta^{\alpha_i\alpha_j}.
\end{equation}
For the fourth order result, we begin by specializing Eq.~\eqref{EQ:Ochiijmatrixelement}
to the case $\hat{O} = \chi^{0\dagger}_{kl}\chi^{0}_{kl}$, which gives
\begin{multline}
\frac{\langle Q \lvert \chi^{0\dagger}_{ij}\chi^{0}_{ij}\chi^{0\dagger}_{kl}\chi^{0}_{kl} \rvert P\rangle}
{\langle Q | P \rangle}
= \frac{1}{16}\bigl( 1 - \epsilon_{ij} 3\delta^{\alpha'_i\alpha'_j}\bigr)\bigl( 1 - \epsilon_{kl} 3\delta^{\alpha_k\alpha_l}\bigr)\\
+ \frac{3}{16}(1+\epsilon_{ij})\epsilon_{kl} \bigl( \delta^{\alpha'_i\alpha'_j} - \delta^{\alpha_i\alpha_j} \bigr).
\end{multline}
Here, $\alpha$ labels the loops in $\bar{Q}P$ and $\alpha'$ labels the
loops in $\bar{Q}P'$ where $P'$ is the modified configuration after
$\chi^{0\dagger}_{kl}\chi^{0}_{kl}$ has acted on $\lvert P \rangle$.
Making use of
\begin{multline} \label{EQ:chi4toS4}
\chi^{0\dagger}_{ij}\chi^{0}_{ij}\chi^{0\dagger}_{kl}\chi^{0}_{kl}
= \frac{1}{16} - \frac{1}{4}\mathbf{S}_i\cdot\mathbf{S}_j
- \frac{1}{4}\mathbf{S}_k\cdot\mathbf{S}_l\\
+ (\mathbf{S}_i\cdot\mathbf{S}_j)(\mathbf{S}_k\cdot\mathbf{S}_l)
\end{multline}
and Eq.~\eqref{EQ:SidotSj}, we arrive at
\begin{multline} \label{EQ:SidotSjSkdotSlalphaprime}
\bigl[ \bigl(\mathbf{S}_i\cdot\mathbf{S}_j\bigr)\bigl(\mathbf{S}_k\cdot\mathbf{S}_l\bigr)\bigr]_{\mathcal{L}}\\
= \epsilon_{ij}\epsilon_{kl} \biggl[\frac{3}{16}\bigl( \delta^{\alpha'_i\alpha'_j} - \delta^{\alpha_i\alpha_j} \bigr)
+ \frac{9}{16} \delta^{\alpha'_i\alpha'_j} \delta^{\alpha_k\alpha_l} \biggr].
\end{multline}
There is no contribution when $\delta^{\alpha_i'\alpha_j'} = \delta^{\alpha_i\alpha_j}=0$.
Hence, the estimator is nonzero only if all four site indices belong 
to the same loop or if there are two indices in each of two loops. The possible 
configurations are shown in Fig.~\ref{FIG:Cijkl}.
Note that if the vertices $(i,j)$ and $(k,l)$ reside in different loops 
or if they are in the same loop but remain unlinked (\emph{i.e.}, there is a path
along the loop from $i$ to $j$ that does not encounter $k$ or $l$) then
$\delta^{\alpha_i'\alpha_j'} = \delta^{\alpha_i\alpha_j}$. 
This equality holds because the operation on $(k,l)$
does not disrupt the loop structure at sites $i$ and $j$.
In this case, the bracketed term on the right-hand side is $\frac{9}{16} \delta^{\alpha_i \alpha_j }$.
Otherwise, $\delta^{\alpha_i'\alpha_j'} = 1-\delta^{\alpha_i\alpha_j}$
 and the bracketed term is $ \frac{3}{16} (1-2\delta^{\alpha_i \alpha_j })$.

The undesirable aspect of Eq.~\eqref{EQ:SidotSjSkdotSlalphaprime} is that it is history-dependent. 
There are several cases to consider because $\alpha'$ references the loop configuration
as it exists after application of the $(k,l)$ vertex. Ideally, we want to reexpress the estimator
in a way that eliminates the $\alpha'$ labels. This can be accomplished---at the expense
of introducing a new quantity $A$---as follows:
\begin{multline} \label{EQ:SidotSjSkdotSl}
\bigl[ \bigl(\mathbf{S}_i\cdot\mathbf{S}_j\bigr)\bigl(\mathbf{S}_k\cdot\mathbf{S}_l\bigr)\bigr]_{\mathcal{L}}\\
\begin{split}
= \epsilon_{ij}\epsilon_{kl} \biggl[&
 -\frac{3}{8}\bigl(1+2A\bigr)
\delta^{\alpha_i\alpha_j}\delta^{\alpha_j\alpha_k}\delta^{\alpha_k\alpha_l}\\
&+ \frac{9}{16}\delta^{\alpha_i\alpha_j}\delta^{\alpha_k\alpha_l}\\
&+ \frac{3}{16}\bigl(\delta^{\alpha_i\alpha_k}\delta^{\alpha_j\alpha_l}
+ \delta^{\alpha_i\alpha_l}\delta^{\alpha_j\alpha_k}\bigr)\biggr].
\end{split}
\end{multline}
$A = 0,1$ is a topological term whose value is nonzero when
traversal of the loop reveals an antisymmetric permulation of the 
site labels $i,j,k,l$. It distinguishes the Fig.~\ref{FIG:Cijkl}(b) configuration ($A=1$) from 
that of Fig.~\ref{FIG:Cijkl}(a) ($A=0)$. For spin correlations of order four and above, 
the loop estimator is no longer simply a function of the loop labels $\alpha_i$.

We can make use of Eqs.~\eqref{EQ:SidotSj} and \eqref{EQ:SidotSjSkdotSl}
to calculate powers of $\hat{\mathbf{M}} = \sum_{i\in A} \mathbf{S}_i - \sum_{j\in B} \mathbf{S}_j$,
which on a bipartite lattice corresponds to the staggered magnetization.
At second order, we have $\hat{\mathbf{M}}^2 = \sum_{ij}\epsilon_{ij} \mathbf{S}_i\cdot
\mathbf{S}_j$ and hence, via
Eq.~\eqref{EQ:SidotSj}, 
\begin{equation} \label{EQ:M2Lalpha}
\mathbf{M}^2_{\mathcal{L}} = \frac{\langle Q \lvert \hat{\mathbf{M}}^2 \rvert P \rangle}{\langle Q | P \rangle}
= \frac{3}{4}\sum_{\alpha=1}^{N_{\circlearrowleft}} L_{\alpha}^2.
\end{equation}
This follows because 
there are $L_{\alpha}^2$ ways to choose two sites from a loop of length
$L_{\alpha}$.
In the same way,
$\hat{\mathbf{M}}^4
= \sum_{ijkl} \epsilon_{ij}\epsilon_{kl}
\bigl(\mathbf{S}_i\cdot\mathbf{S}_j\bigr)\bigl(\mathbf{S}_k\cdot\mathbf{S}_l\bigr)$
can be computed using Eq.~\eqref{EQ:SidotSjSkdotSl}.
The only complication is the case where $i,j,k,l$ are all in the same loop $\alpha$.
Of the $L_\alpha^4$ such configurations, 
\begin{equation} \label{EQ:Aeq1count}
L_{\alpha}\sum_{j=1}^{L_\alpha} \bigl[ 2j(L_\alpha-j)-1 \bigr] = \frac{1}{3}L_\alpha^4 - \frac{4}{3}L_\alpha^2
\end{equation}
have weight $-3/16$. The counting reflects the fact that there are $L_\alpha$ ways to fix $i$ and, 
as $j$ ranges over the loop, $2j(L_\alpha-j)-1$ ways to place $k$ and $l$ in opposite loop segments.
(This includes mixed cases of the form $i=k$, $j\neq l$ which are $A=1$ in nature,
but omits those of the form $i=k$, $j=l$, which are $A=0$.)
The remainder have weight $9/16$.
Thus, the total contribution is
\begin{multline}
\sum_\alpha \biggl[ \frac{9}{16}\biggl(\frac{2}{3}L_\alpha^4 + \frac{4}{3}L_\alpha^2
\biggr) - \frac{3}{16}\biggl(\frac{1}{3}L_\alpha^4 - \frac{4}{3}L_\alpha^2\biggr) \biggr]\\
+ \biggl(2\times\frac{3}{16}+\frac{9}{16}\biggr)\sum_{\alpha\neq\beta} L_\alpha^2L_\beta^2,
\end{multline}
which simplifies to
\begin{equation}
\begin{split} \label{EQ:M4Lalpha}
\mathbf{M}^4_{\mathcal{L}}
&= \sum_\alpha \biggl( -\frac{5}{8}L_\alpha^4 + L_\alpha^2
\biggr) + \frac{15}{16}\biggl(\sum_\alpha L_\alpha^2\biggr)^2.
\end{split}
\end{equation}

\section{\label{SEC:Cumulant} Cumulant generating function}

Calculating spin correlation functions as we did in the previous section
involves keeping track of all possible ways that a given number of
site index pairs, $(i,j),(k,l),\ldots,$ can be assigned to a set of valence bond loops. 
For each such assignment, we then have to determine the contribution 
to the correlation function based on how the loops are reconfigured
by the valence bond operators $\chi^{0\dagger}_{ij} \chi^{0}_{ij}, \chi^{0\dagger}_{kl} \chi^{0}_{kl},
\ldots.$
This brute force approach 
becomes increasingly 
cumbersome (and errorprone)
as the order of the correlation 
function becomes large. At second and fourth order, the problem is manageable.
For $\mathbf{S}_i \cdot \mathbf{S}_j$, there are are two distinct configurations---one 
that contributes (when $i$ and $j$ are in the same loop) and one that
does not (when $i$ and $j$ are in different loops)---and for
$(\mathbf{S}_i \cdot \mathbf{S}_j)(\mathbf{S}_j \cdot \mathbf{S}_l)$,
there are eight configurations---four that contribute (those shown in Fig.~\ref{FIG:Cijkl}) 
and four that do not. But at sixth order, there are already 33 configurations to consider---an 
intractable nightmare.

In this section, we present a vastly more simple approach that emerges from a deeper 
understanding of the relationship between correlation functions and loops.
To start, we introduce a new operator
\begin{equation}
\hat{\gamma}_{ij} = \frac{1}{4} + \epsilon_{ij}\mathbf{S}_i\cdot\mathbf{S}_j
= \frac{1}{4}\bigl(1+\epsilon_{ij}\bigr) - \epsilon_{ij} \chi^{0\dagger}_{ij}
\chi^{0}_{ij},
\end{equation}
which is designed to have a positive definite matrix element, irrespective of the
sublattice membership of $i$ and $j$. 
What motivates this definition is a desire to compensate for the asymmetry between
Eq.~\eqref{EQ:updateinjm} and Eqs.~\eqref{EQ:updateinim} and \eqref{EQ:updatejnjm},
which describe the effect of $\chi^{0\dagger}_{ij}\chi^{0}_{ij}$ on a valence bond state.
In the AB case ($\epsilon_{ij} = -1$), the operator $\hat{\gamma}_{ij} = \chi^{0\dagger}_{ij}\chi^{0}_{ij}$
produces only a rearrangement of bonds and leaves the phase of the state 
unchanged, which is the desired state of affairs;
in the AA/BB case ($\epsilon_{ij} = 1$), the operator 
$\hat{\gamma}_{ij} = \tfrac{1}{2}-\chi^{0\dagger}_{ij} \chi^{0}_{ij}$
is engineered to behave in exactly the same way, by cancelling the first term
and changing the sign of the second term on the right-hand side of 
Eqs.~\eqref{EQ:updateinim} and \eqref{EQ:updatejnjm}.
Pictorially, this amounts to removing the identity part 
and changing the sign of the
transposition part from $-\tfrac{1}{2}$ to $\tfrac{1}{2}$
on the right-hand side of the third (bottom) update rule in Fig.~\ref{FIG:updates}.
The rearrangement of individual singlets is governed by
\begin{align}
\hat{\gamma}_{ij}[i,j] &= \hat{\gamma}_{ii}[i,j] = [i,j] \\
\hat{\gamma}_{ij}[i,l][k,j] &= \hat{\gamma}_{ik}[i,l][k,j] = \frac{1}{2}[i,j][k,l]
\end{align}
for sites $i,k \in A$ and $j,l \in B$. Hence, 
\begin{align} \label{EQ:gammainjnP}
\hat{\gamma}_{i_nj_m} \lvert P \rangle &= 2^{\delta_{Pn,m}-1}\lvert (Pn\ m) P \rangle,\\
\hat{\gamma}_{i_ni_m} \lvert P \rangle &= 2^{\delta_{n,m}-1}\lvert (Pn\ Pm) P \rangle,\\
\hat{\gamma}_{j_nj_m} \lvert P \rangle &= 2^{\delta_{n,m}-1}\lvert (n\ m) P \rangle.
\end{align}

\begin{figure}
\begin{center}
\includegraphics{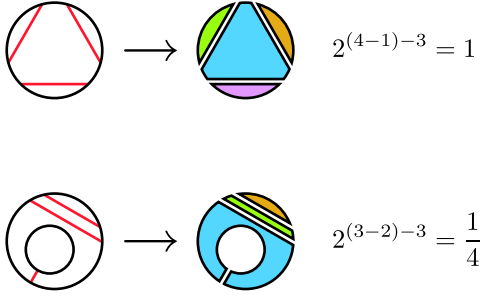}
\end{center}
\caption{ \label{FIG:Cijklmn}
Two examples of the correlation function defined in Eq.~\eqref{EQ:correlation-gamma}
acting at third order.}
\end{figure}

Since $\hat{\gamma}_{ij}$ involves only the transposition of bond pairs,
its effect on loops is limited to splitting one loop into two (when $i$ and $j$ belong
to the same loop) and joining two loops into one (when $i$ and $j$ belong to different loops).
Its correlation function obeys the simple rule
\begin{equation} \label{EQ:correlation-gamma}
C_{\underbrace{ijkl\cdots}_{\text{$2n$ indices} }} 
= \frac{ \langle Q \lvert \hat{\gamma}_{ij}\hat{\gamma}_{kl}\cdots \rvert P \rangle }
{ \langle Q | P \rangle }
= 2^{\Delta N_{\circlearrowleft} - n} > 0,
\end{equation}
where $\Delta N_{\circlearrowleft}$ represents the change in loop number
after the loops have been split or joined at ($i,j$), ($k,l$), \emph{etc}.
Two examples are given in Fig.~\ref{FIG:Cijklmn}.

\begin{figure*}
\begin{center}
\includegraphics{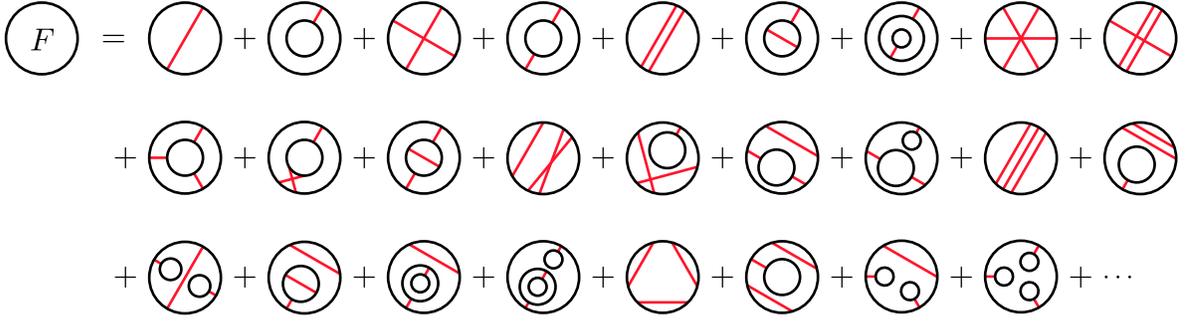}
\end{center}
\caption{ \label{FIG:correlations1}
The diagrammatic expansion of the cumulant generating function 
$F$ [from Eq.~\eqref{EQ:Fgenerating}] is shown to third order in the coupling $a_{ij}$.
The black lines denote valence bond loops and the red lines $\hat{\gamma}$ operators.
}
\end{figure*}

\begin{figure}
\begin{center}
\includegraphics{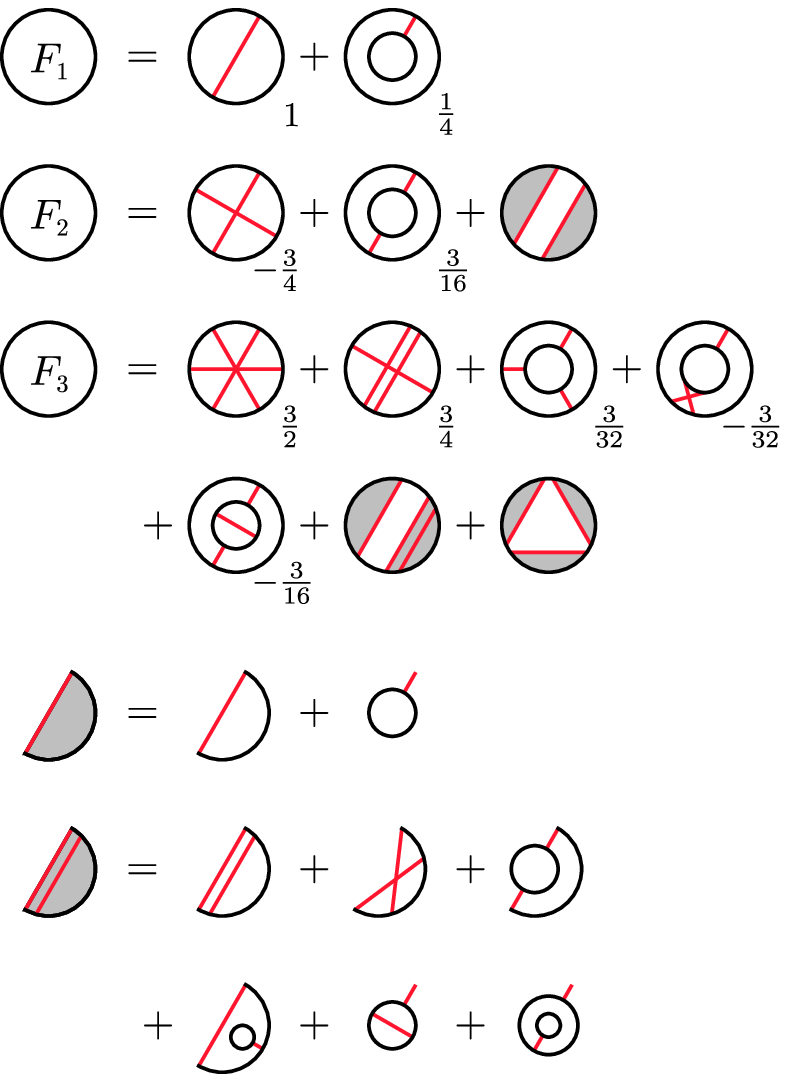}
\end{center}
\caption{ \label{FIG:correlations2}
At each order, the contributions to $F$ can be classified according to their irreducibility.
Nonirreducible diagrams are those that can be stitched together from self-energy parts of
lower order in $\gamma$. The self-energy parts, shown as shaded arcs, are built
by removing a single loop segment from the diagrams in Fig.~\ref{FIG:correlations1}. 
The values listed next to each diagram represent the $\tilde{C}$ cumulant value
associated with that configuration. $\tilde{C} = 0$ for the composite terms.
}
\end{figure}

We now introduce a generating function
\begin{equation} \label{EQ:Fgenerating}
F[a] = \log \langle Q \lvert \exp \biggl( \sum_{ij} a_{ij} \hat{\gamma}_{ij} \biggr) \rvert P \rangle.
\end{equation}
Its diagramatic expansion in powers of the coupling $a_{ij}$, shown in Fig.~\ref{FIG:correlations1},
is completely analogous to the Goldstone diagrams familiar from standard
many-body theory. (There are only connected diagrams because of the linked-cluster theorem.)
The $n^{\text{th}}$ order derivatives of $F$ are the cumulants of the correlators defined in
Eq.~\eqref{EQ:correlation-gamma},
\begin{equation}
\tilde{C}_{\underbrace{ijk\cdots}_{\text{$2n$ indices} }} =  \frac{ \partial^n F}{\partial a_{ij}
\partial a_{kl} \cdots }\biggr\rvert_{a=0},
\end{equation}
and are related to them by
\begin{align}
\tilde{C}_{ij} &= C_{ij}, \\
\tilde{C}_{ijkl} &= C_{ijkl} - \tilde{C}_{ij}\tilde{C}_{kl}, \\
\begin{split}
\tilde{C}_{ijklmn} &= C_{ijklmn} - \tilde{C}_{ij}\tilde{C}_{klmn} \\
                               &\quad - \tilde{C}_{kl}\tilde{C}_{ijmn} - \tilde{C}_{mn}\tilde{C}_{ijkl}\\
                                &\quad - \tilde{C}_{ij}\tilde{C}_{kl}\tilde{C}_{mn}.
\end{split}
\end{align}
What is important about the cumulants is that their subtractive terms lead to perfect
cancellation (\emph{e.g.}, $C_{ijkl} = C_{ij}C_{kl}$ so that $\tilde{C}_{ijkl} = 0$) 
whenever the $\hat{\gamma}$-vertices connect the loops into a network that 
cannot be disentangled by cutting it along two loop segments.
The exact statement is that a cumulant vanishes unless its
configuration is irreducible in the standard many-body diagram sense.
See Fig.~\ref{FIG:correlations2}.

The cumulants, in turn, are related to the physical spin correlation functions by
\begin{align} \label{EQ:CijtoS}
\tilde{C}_{ij} &= S_{ijkl}+\tfrac{1}{4} \\ \label{EQ:CijkltoS}
\tilde{C}_{ijkl} &= S_{ijkl} - S_{ij}S_{kl} \\ \label{EQ:CijklmntoS}
\begin{split}
\tilde{C}_{ijklmn} &= S_{ijklmn} - S_{ij}S_{klmn} \\
                               &\quad - S_{kl}S_{ijmn} - S_{mn}S_{ijkl}\\
                                &\quad + 2S_{ij}S_{kl}S_{mn}.
\end{split}
\end{align}
Here, we have used the shorthand 
\begin{equation} \label{EQ:Sshorthand}
S_{ijkl\cdots} = (\epsilon_{ij}\epsilon_{kl}\cdots)\frac{\langle Q \lvert (\mathbf{S}_i\cdot\mathbf{S}_j)
 (\mathbf{S}_k\cdot\mathbf{S}_l) \cdots
\rvert P \rangle}{\langle Q | P \rangle}.
\end{equation}
Thus, computing spin correlation functions at any order is just a matter of determining
the relevant irreducible diagrams, computing the corresponding cumulants, and 
then solving for the desired term in Eqs.~\eqref{EQ:CijtoS}--\eqref{EQ:CijklmntoS}.
By this means, we recover the loop estimators Eq.~\eqref{EQ:SidotSj}
and Eq.~\eqref{EQ:SidotSjSkdotSl}.

It is also possible to compute even powers of the staggered magnetization by
summing over all the site indices:
\begin{align}
\sum_{ij} \tilde{C}_{ij} &= \mathbf{M}^2_{\mathcal{L}} + N^2,\\
\sum_{ijkl} \tilde{C}_{ijkl} &= \mathbf{M}^4_{\mathcal{L}} - \bigl( \mathbf{M}^2_{\mathcal{L}} \bigr)^2, \\
\sum_{ijklmn}\!\! \tilde{C}_{ijklmn} &= \mathbf{M}^6_{\mathcal{L}}
- 3 \mathbf{M}^2_{\mathcal{L}} \mathbf{M}^4_{\mathcal{L}} + 2\bigl( \mathbf{M}^2_{\mathcal{L}} \bigr)^3.
\end{align}
At second order in $\hat{\mathbf{M}}$ (first order in $\hat{\gamma}_{ij}$), there
are two irreducible diagrams: a one-loop configuration of weight 1 and a two-loop
configuration of weight 1/4. Since there are $L_{\alpha}^2$ ways to select two sites in
loop $\alpha$ and $L_{\alpha}L_{\beta}$ ways to select one site in loop $\alpha$ 
and one in loop $\beta$, we have
\begin{equation}
\begin{split}
\mathbf{M}^2_{\mathcal{L}} &= 1\sum_\alpha L_\alpha^2 + \frac{1}{4} \sum_{\alpha\neq\beta} L_\alpha L_\beta - N^2 \\
&= \frac{3}{4}\sum_{\alpha} L_\alpha^2 + \frac{1}{4} \biggl( \sum_\alpha L_\alpha \biggr)^2
- N^2 \\
&= \frac{3}{4}\sum_{\alpha} L_\alpha^2.
\end{split}
\end{equation}
A glance at Eq.~\eqref{EQ:M2Lalpha} confirms that this method produces the correct result.

At fourth order in $\hat{\mathbf{M}}$ (second order in $\hat{\gamma}_{ij}$), there are again
only two irreducible diagrams: a one-loop (cross) configuration of weight $-3/4$ and a two-loop
configuration of weight $3/16$. We have already worked out the counting 
in Eq.~\eqref{EQ:Aeq1count}. Hence,
\begin{equation}
\begin{split}
\mathbf{M}^4_{\mathcal{L}} &= -\frac{3}{4}\sum_\alpha \Bigl(\frac{1}{3}L_\alpha^4 - \frac{4}{3}L^2_\alpha\Bigr) + \frac{3}{16} \sum_{\alpha\neq\beta} 2L_\alpha^2 L_\beta^2 
+\bigl(\mathbf{M}^2_{\mathcal{L}}\bigr)^2 \\
&= \sum_{\alpha} \Bigl(-\frac{5}{8}L_\alpha^4 + L_{\alpha}^2\Bigr) + \frac{3}{8} \biggl( \sum_\alpha L_\alpha^2 \biggr)^2 +  \biggl(\frac{3}{4} \sum_\alpha L_\alpha^2 \biggr)^2 \\
&= \sum_{\alpha} \Bigl(-\frac{5}{8}L_\alpha^4 + L_{\alpha}^2\Bigr) + \frac{15}{16} \biggl( \sum_\alpha L_\alpha^2 \biggr)^2,
\end{split}
\end{equation}
which is the result of Eq.~\eqref{EQ:M4Lalpha}. 

At sixth order in $\hat{\mathbf{M}}$ (third order in $\hat{\gamma}_{ij}$), there are five
irreducible diagrams. The diagrams of weight $3/32$ and $-3/32$ both number
$\sum_{\alpha\neq\beta}2L_{\alpha}^3L_{\beta}^3$ and thus cancel each other,
so we only need to consider the other three: a one-loop (asterisk) configuration
of weight $3/2$, a one-loop (railway tie) configuration of weight $3/4$, and
a two-loop configuration of weight $-3/16$. Counting the number of site index 
arrangements for these three cases requires some work, but it is no different in principle from 
what we did in Eq.~\eqref{EQ:Aeq1count}. We skip the details and simply
report the result:
\begin{equation}
\begin{split}
\mathbf{M}^6_{\mathcal{L}}
&= \frac{3}{2}\sum_{\alpha} \Bigl( \frac{1}{15}L_{\alpha}^6 - \frac{16}{15}L_{\alpha}^2\Bigr)\\
&\quad+ \frac{3}{4}\sum_{\alpha} \Bigl( \frac{1}{5}L_{\alpha}^6 - \frac{4}{3}L_{\alpha}^4 
+ \frac{32}{15}L_{\alpha}^2\Bigr)\\
&\quad- \frac{3}{16} 2\sum_{\alpha\neq\beta} L_{\alpha}^2L_{\beta}^2\bigl(
L_{\alpha}^2+L_{\beta}^2-4\bigr)\\
&\quad+ 3\mathbf{M}^2_{\mathcal{L}}\mathbf{M}^4_{\mathcal{L}} - 2\bigl(\mathbf{M}^2_{\mathcal{L}}
\bigr)^3.
\end{split}
\end{equation}
This simplifies to
\begin{equation}
\begin{split}
\mathbf{M}^6_{\mathcal{L}}
&= \sum_{\alpha} \Bigl(L_{\alpha}^6 - \frac{5}{2}L_{\alpha}^4\Bigr)
+\frac{81}{64}\biggl( \sum_\alpha L_\alpha^2 \biggr)^3\\
&\quad -\frac{69}{32}\biggl( \sum_\alpha L_\alpha^4 \biggr)\biggl( \sum_\alpha L_\alpha^2 \biggr).
\end{split}
\end{equation}

In terms of the $\bar{Q}P$ cycle lengths ($L_\alpha \to 2k_\alpha$), the 
magnetization formulas are
\begin{align} \label{EQ:M2kalpha}
\mathbf{M}^2_{\mathcal{L}} &= 3\sum_{\alpha} k^2_{\alpha}, \\ \label{EQ:M4kalpha}
\mathbf{M}^4_{\mathcal{L}} &= 
\sum_{\alpha} \bigl(-10k^4_{\alpha}+4k^2_{\alpha}\bigr)
+15 \biggl(\sum_{\alpha} k^2_{\alpha}\biggr)^2, \\
\begin{split}
\mathbf{M}^6_{\mathcal{L}}  &= \sum_{\alpha} \bigl(64k^6_{\alpha}-40k^4_{\alpha}\bigr)
+ 81\biggl(\sum_{\alpha} k^2_{\alpha}\biggr)^3\\
&\quad- 138 \biggl(\sum_{\alpha} k^4_{\alpha}\biggr)\biggl(\sum_{\alpha} k^2_{\alpha}\biggr).
\end{split}
\end{align}
Recall that in Sect.~\ref{SEC:Overlaps} we used 
$n_k = \sum_{\alpha=1}^{N_{\circlearrowleft}}\delta(k-k_{\alpha})$,
the cycle length distribution function, to derive an expression for the overlap $\langle Q | P \rangle$.
It is easy to see that the second moment of $n_k$ is related to the presence
of long-range antiferromagnetic order in the system:
a loop average of Eq.~\eqref{EQ:M2kalpha} gives
$\langle \hat{\mathbf{M}}^2\rangle = \sum_k 3k^2 \langle n_k \rangle_W$.
Similarly, the fourth and sixth powers of the staggered magnetization have terms
involving the cycle-length correlations $\langle n_kn_{k'} \rangle_W$ and
$\langle n_kn_{k'}n_{k''} \rangle_W$.

\begin{figure}
\begin{center}
\includegraphics{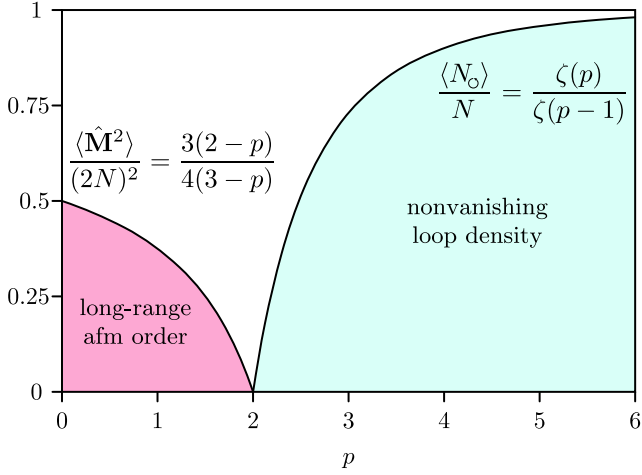}
\end{center}
\caption{ \label{FIG:lrorder}
The squared staggered magnetization and the loop density are plotted 
as a function of the exponent $p$
for a cycle-length distribution of the form $\langle n_k \rangle_W \sim k^{-p}$.
The two quantities are complementary indicators of the long range order below $p=2$.
 }
\end{figure}

The key feature is the tail of the cycle-length distribution. If it is too weak, then the
loop gas is characterized by a macroscopic number of small loops.
In the opposite limit, the system has a vanishing number density of 
loops, although each loop contains a nonvanishing fraction of all the spins.
These system-spanning loops are the foundation of the long-range order.
As an example, suppose that $\langle n_k \rangle_W \sim k^{-p}$, with the 
normalization set by the constraint $N = \sum_{k=1}^N k\langle n_k \rangle_W$.
The standard large-$N$ summation rules are
\begin{equation}
\sum_{k=1}^N k^{-p} = \begin{cases}
\frac{1}{1-p}N^{1-p} + \mathcal{O}(N^{-p}) & \text{if $p<1$} \\
\log N + \mathcal{O}(1) & \text{if $p=1$}\\
\zeta(p) + \mathcal{O}(N^{1-p}) & \text{if $p > 1$},
\end{cases}
\end{equation}
where $\zeta(p)$ is the Reimann Zeta function.
These can be used to derive the asymptotic behaviour listed in Table~\ref{TAB:thermlimit}.
Figure~\ref{FIG:lrorder} depicts the $N\to\infty$ behaviour of the
loop number density and the average staggered moment:
\begin{align}
\frac{\langle N_{\circlearrowleft} \rangle}{N} &= \frac{1}{N} \sum_{k=1}^N \langle n_k \rangle_W 
\to \frac{\zeta(p)}{\zeta(p-1)}\theta(p-2), \\
\frac{\langle \hat{\mathbf{M}}^2 \rangle}{(2N)^2} &= \frac{3}{4N^2}\sum_{k=1}^N k^2 \langle n_k \rangle_W
\to \frac{3(2-p)}{4(3-p)}\theta(2-p).
\end{align}
Clearly, long-range order exists for a range of exponents $0 \le p < 2$.
The phase transition at the critical value $p_{\text{c}}=2$ should be visible as a sharp step
in the function $\langle \hat{\mathbf{M}}^2 \rangle/\langle N_{\circlearrowleft}\rangle^2$
and in the Binder ratios~\cite{Binder81} $\langle \hat{\mathbf{M}}^4 \rangle/\langle \hat{\mathbf{M}}^2\rangle^2$
and $\langle \hat{\mathbf{M}}^6 \rangle/\langle \hat{\mathbf{M}}^2\rangle^3$.

\begin{table}
 \caption{\label{TAB:thermlimit} Asymptotic behaviour of the loop density and staggered magnetization 
 in the thermodynamic limit as a function of the loop distribution tail.}
\begin{ruledtabular}
\begin{tabular}{@{\quad}c|c|c@{\quad}}
$\langle n_k \rangle_W \sim k^{-p}$ & $\mathcal{O}(\langle N_{\circlearrowleft} \rangle/N)$ & 
$\mathcal{O}(\langle \hat{\mathbf{M}}^2 \rangle/N^2)$  \tabularnewline \hline
$0 \le p < 1$ & $\frac{1}{N}$ & $1$ \tabularnewline
$p = 1$ & $\frac{\log N}{N}$ & $1$ \tabularnewline
$1 < p < 2$ & $\frac{1}{N^{2-p}}$ & $1$ \tabularnewline
$p = 2$ & $\frac{1}{\log N}$ & $\frac{1}{\log N}$ \tabularnewline
$2 < p < 3$ & $1$ & $\frac{1}{N^{p-2}}$ \tabularnewline
$p \ge 3$ & $1$ & $\frac{1}{N}$ \tabularnewline
short-ranged & $1$ & $\frac{1}{N}$
\end{tabular}
\end{ruledtabular}
\end{table}

In the $p\to \infty$ limit, the cycle-length distribution becomes sharply peaked 
at the minimum length, $\langle n_k \rangle_W = N\delta_{k,1}$, which is the result 
for a fixed dimer configuration.
More generally, a generic VBS state will have $\langle n_k \rangle_W 
\sim \theta(N_0-k)$, where $N_0$ is the size of the plaquet (the basic unit 
of translational symmetry breaking). When the distribution is short-ranged---having
either an upper cutoff or an exponentially suppressed tail---there is no magnetic order.

\section{\label{SEC:Extended} Valence bond coverings of the full Hilbert space}

In previous sections, we discussed the properties of the $S=0$ valence bond states.
We now turn our attention to the full Hilbert space and consider 
\emph{extended} valence bond states of the form
\begin{equation} 
\lvert P; \vec{\mu} \rangle = \prod_{n=1}^N\chi^{\mu_n\dagger}_{i_n,Pj_n}\lvert \text{vac} \rangle.
\end{equation}
Here, $\vec{\mu} = (\mu_1,\mu_2, \ldots, \mu_N) \in \mathbb{Z}_4^N$
is a vector of indices describing the singlet/triplet character of each bond.
The set of such states is overcomplete with respect to the full Hilbert space
($4^NN! \gg 2^{2N}$).
The restriction to AB bonds is still meaningful in the extended basis 
since for every index pair $\mu'',\nu''$ there is at least one set of indices $\mu,\nu,\mu',\nu'$
(in fact there are exactly four) such that the equation
\begin{equation}
c_1 \chi^{\mu\dagger}_{il}\chi^{\nu\dagger}_{jk}
+ c_2 \chi^{\mu'\dagger}_{ij}\chi^{\nu'\dagger}_{kl}
+ c_3 \chi^{\mu''\dagger}_{lj}\chi^{\nu''\dagger}_{ik} = 0
\end{equation}
has a nontrivial solution. This is the singlet/triplet generalization of Eq.~\eqref{EQ:linearly-dependent}.

\begin{figure}
\begin{center}
\includegraphics{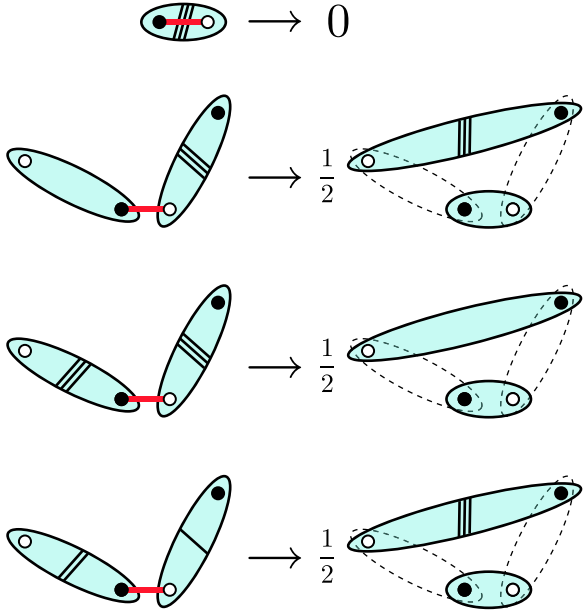}
\end{center}
\caption{ \label{FIG:updates2}
Summary of the update rules for triplet bonds.
The number of cross hatches on each bond indicates its triplet character.
 }
\end{figure}

To understand how triplet states evolve under the Heisenberg hamiltonian, 
we must reproduce the analysis of Sec.~\ref{SEC:Evolution} but now with
Eqs.~\eqref{EQ:chi2-commutator} and \eqref{EQ:chi3-commutator}
specialized to 
\begin{equation} 
[\chi^{0}_{ij},\chi^{\mu\dagger}_{ij}]\lvert \text{vac} \rangle = \delta^{0\mu}\lvert \text{vac} \rangle
\end{equation}
and
\begin{equation} 
[\chi^{0}_{ij},\chi^{\mu\dagger}_{kj}\chi^{\nu\dagger}_{il}]\lvert \text{vac} \rangle = \frac{1}{2}\,T^{\lambda\mu 0\nu}\,\chi^{\lambda\dagger}_{kl}
\lvert \text{vac} \rangle.
\end{equation}
As before, the operator $-\hat{H}_{ij}$ has both a diagonal and offdiagonal part.
If there is a pre-existing singlet bond across the active sites, the bond configuration
is left unchanged. If there is a pre-existing triplet bond, however, the state is annihilated.
Otherwise, the bonds are rearranged and the singlet/triplet labels reassigned
in a spin-conserving fashion. The update rule for $-\hat{H}_{ij}$ with $i\in A$ and $j\in B$,
illustrated in Fig.~\eqref{FIG:updates2}, is given by
\begin{widetext}
\begin{equation}
-\hat{H}_{i_nj_m} \lvert P;\vec{\mu} \rangle = \begin{cases}
\delta^{\mu_n 0} \lvert P; \vec{\mu} \rangle & \text{if $Pm = n$}\\
\frac{1}{2}T^{\lambda \mu_n 0\mu_{\bar{P}m} }\lvert (Pn\ m) P; (\mu_1,\ldots,
\mu_{n-1},0,\mu_{n+1},\ldots,\mu_{\bar{P}m-1},
\lambda,\mu_{\bar{P}m+1},\ldots,\mu_N)\rangle & \text{otherwise}\\
\end{cases}
\end{equation}
\end{widetext}

There are some important differences from the purely singlet-bond case.
There is now the possibility of anihilating a state
by acting directly on a triplet bond, 
unlike in the singlet sector, where $(\hat{H}_{ij}\hat{H}_{i'j'} \hat{H}_{i''j''}\cdots) \lvert P \rangle \neq 0$
always. Also, when bonds of different species interact,
the weight $T^{\lambda\mu 0\nu}$ is potentially negative:
the $\tau$ operators obey the identity
\begin{equation}
\tau^\mu\tau^{0\dagger}\tau^\nu = \begin{pmatrix}
\tau^0 & \tau^1 & \tau^2 & \tau^3\\ 
\tau^1 & \tau^0 & \tau^3 & -\tau^2\\
\tau^2 & -\tau^3 & \tau^0 & \tau^1\\
\tau^3 & \tau^2 & -\tau^1 & \tau^0
\end{pmatrix}_{\mu\nu}
\end{equation}
and hence
\begin{equation}
T^{\lambda\mu 0 \nu} = \delta^{\lambda 0}\delta^{\mu\nu} +  \epsilon^{\mu\nu\lambda},
\end{equation}
where $\epsilon^{\mu\nu\lambda}$ is the alternating symbol.

Overlaps of extended valence bond states can be computed as before, only now we must use
the most general form of the anticommutation relationship in Eq.~\eqref{EQ:chi3-commutator}.
We find that
\begin{equation} \label{EQ:QvuPmuoverlap}
\langle Q; \vec{\nu} | P; \vec{\mu} \rangle = 
2^{N_{\circlearrowleft}-N}
\prod_{\alpha=1}^{N_{\circlearrowleft}} K_{\alpha}.
\end{equation}
Here, $N_{\circlearrowleft}$ is the number of cycles in $\bar{Q}P$
and $K_{\alpha}$ is a chirality factor whose value is given by the trace of
the product of $\tau^{\nu_n\dagger}\tau^{\mu_n}$ in proper order around each cycle.
The form of $K_{\alpha}$ follows from the
``boxcar'' property
\begin{equation}
\sum_\lambda\tr\bigl( \cdots \tau^{\nu\dagger}\tau^{\lambda}
\bigr)T^{\lambda\mu\nu'\mu'} = \tr\bigl(\cdots\tau^{\nu\dagger}\tau^{\mu}
\tau^{\nu'\dagger}\tau^{\mu'}\bigr).
\end{equation}
For example, the loop $\bar{Q}P = (1\ 2\ 4\ 5\ 3)\cdots$, 
highlighted in Fig.~\ref{FIG:overlap},
has associated with it a factor
\begin{equation}
K_{\alpha} = \frac{1}{2}\tr \bigl(\tau^{\nu_1\dagger}\tau^{\mu_1}
\tau^{\nu_2\dagger}\tau^{\mu_2}
\tau^{\nu_4\dagger}\tau^{\mu_4}
\tau^{\nu_5\dagger}\tau^{\mu_5}
\tau^{\nu_3\dagger}\tau^{\mu_3}\bigr).
\end{equation}

Since $K_{\alpha}$ takes the values $0$ or $\pm 1$, the overlap is no longer
guaranteed to be positive definite, as it was in the case of only singlet bonds. 
If we do not mix triplet species, however, positive definiteness is preserved.
In that case, $K_{\alpha} = 1$ if the loop contains an even number of triplets
and $K_{\alpha} = 0$ if the loop contains an odd number. More generally,
if we move around a given loop in proper order, encountering triplets
of species $a,b,c,\ldots$ along the way, then $K_{\alpha}$ takes the value
\begin{equation}
\begin{split}
0 &\quad \text{one triplet} \\
(-1)^{1+|\vec{\mu}|}\delta^{ab} &\quad \text{two triplets} \\
(-1)^{1+|\vec{\mu}|}\epsilon^{abc} &\quad \text{three triplets} \\
(-1)^{|\vec{\mu}|}(\delta^{ab}\delta^{cd} + \delta^{ad}\delta^{bc} - \delta^{ac}\delta^{bd})
&\quad \text{four triplets}
\end{split}
\end{equation}
where $|\vec{\mu}|$ counts the number of nonzero entries (triplets) 
in $\vec{\mu}$ that belong to the loop in question.
In other words, the overall sign depends on how many triplets arise
from $\lvert P; \vec{\mu} \rangle$ and how many from $\lvert Q; \vec{\nu} \rangle$.

\section{\label{SEC:Components} Staggered Magnetization Components}

The staggered magnetization operator has the special property that it can always be 
written so as to induce no rearrangement of bonds. This is simply a matter of grouping its
terms to match the particular bond tiling of the state it is acting on.
That is, for any $P \in \mathcal{S}_N$,
\begin{equation} \label{EQ:Manorearrange}
\hat{M}^a = \sum_{i\in A} S_i^a - \sum_{j \in B} S_j^a
= \sum_{n=1}^N \bigl(S_{i_n}^a - S_{j_{Pn}}^a\bigr).
\end{equation}
Since the bracketed term behaves according to Eq.~\eqref{EQ:SiminusSj},
we see that the effect of $\hat{M}^a$ on an $S=0$ valence bond state is to create
a superposition of states, each of which has one bond promoted to an $a$-triplet:
\begin{equation}
\hat{M}^a \lvert P \rangle = i \sum_{n=1}^N \lvert P; a_n \rangle.
\end{equation}
Here, $\lvert P; a_n \rangle$ denotes the state $\lvert P; \vec{\mu} \rangle$
with $\vec{\mu}$ having all zero entries except for $a$ in the $n^{\text{th}}$ slot.
Since lone triplets always produce zero-contribution loops, we find that
\begin{equation}
M^a_{\mathcal{L}} = \frac{\langle Q \rvert \hat{M}^a \lvert P \rangle}{\langle Q | P \rangle}
 =  i \sum_{n=1}^N \frac{\langle Q | P; a_n \rangle}{\langle Q | P \rangle} = 0,
\end{equation}
which is the expected result for rotationally invariant states $\lvert P \rangle$
and $\lvert Q \rangle$.
Two triplets of the same species, however, if they originate in different states,
give unit weight whenever they lie in the same
loop. Thus,
\begin{equation} \label{EQ:MaMb}
(M^aM^b)_{\mathcal{L}}
 = \sum_{m=1}^N \sum_{n=1}^N \frac{\langle Q; a_m | P; b_n \rangle}{\langle Q | P \rangle}
 = \delta^{ab}\sum_{\alpha} k_{\alpha}^2.
\end{equation}
The final equality holds because, in a cycle of length $k_\alpha$, there are $k_\alpha$ ways 
to place the $m^{\text{th}}$ $Q$ link and $k_\alpha$ ways to place the $n^{\text{th}}$ $P$ link.

Correlation functions at fourth order can be computed in much the same way.
Two applications of $\hat{M}^a$ to a valence bond state yields
\begin{equation}
(\hat{M}^a)^2\lvert P \rangle 
= N\lvert P \rangle - \sum_{m\neq n} \lvert P; a_m,a_n \rangle,
\end{equation}
and the overlap of two such states is
\begin{widetext}
\begin{equation} \label{EQ:Ma2Mb2uneval}
\frac{\langle Q \lvert (\hat{M}^a)^2(\hat{M}^b)^2 \rvert P \rangle}{\langle Q | P \rangle}
= N^2 - \sum_{m\neq n}\biggl( \frac{\langle Q; a_m,a_n | P \rangle}{\langle Q | P \rangle}
+ \frac{\langle Q | P; b_m,b_n \rangle}{\langle Q | P \rangle}\biggr)
 + \sum_{m\neq n}\sum_{m'\neq n'} 
\frac{\langle Q; a_m,a_n | P; b_{m'},b_{n'} \rangle}{\langle Q | P \rangle}.
\end{equation}
\end{widetext}

The bracketed terms in Eq.~\eqref{EQ:Ma2Mb2uneval} each take the
value $-1$ if the two triplets are in the same loop and vanish otherwise.
In the case of one triplet species ($a=b$), the final term contributes if there is either
a single loop containing all four triplets or a pair of loops containing 
two triplets each. These configurations all have weight +1.
Applying the appropriate counting arguments yields
\begin{multline} \label{EQ:Ma4count}
(M^a)^4_{\mathcal{L}} = N^2 + 2N\sum_{\alpha}k_\alpha(k_\alpha-1)
+ \sum_\alpha k_\alpha^2(k_\alpha-1)^2\\
+ \sum_{\alpha\neq\beta}k_\alpha(k_\alpha-1)k_\beta(k_\beta-1)
+ 2\sum_{\alpha\neq\beta}k_\alpha^2k_\beta^2.
\end{multline}
Note that the $m\neq n$ and $m' \neq n'$ constraints have no effect when
$m$ and $m'$ ($m$ and $n'$) are in one cycle and $n$ and $n'$ ($n$ and $m'$) 
are in another. Equation~\eqref{EQ:Ma4count}
simplifies considerably to give
\begin{equation} \label{EQ:Ma4}
(M^a)^4_{\mathcal{L}}  = -2\sum_\alpha k_\alpha^4 + 3\biggl(\sum_{\alpha} k_\alpha^2\biggr)^2.
\end{equation}

If there are two triplet species $(a\neq b)$, the situation is slightly more complicated.
When all four triplets are in the same loop, there are 
\begin{equation} \label{EQ:4tripletsw1}
2k_\alpha \sum_{n=3}^{k_\alpha} (n-1)(n-2) = \frac{2}{3}k_\alpha^4 - 2k_\alpha^3 + \frac{4}{3}k_\alpha^2 
\end{equation}
configurations, having weight $+1$, in which the $a$ and $b$ triplets
appear in consecutive order around the loop ($aabb$) and
\begin{equation} \label{EQ:4tripletswm1}
2k_\alpha \sum_{n=2}^{k_\alpha} (n-1)(k-n+1) = -\frac{1}{3}k_\alpha^2 + \frac{1}{3}k_\alpha^4
\end{equation}
configurations, having weight $-1$, in which they appear in alternating order ($abab$).
[Note that Eqs.~\eqref{EQ:4tripletsw1} and \eqref{EQ:4tripletswm1} sum to 
$k_\alpha^2(k_\alpha-1)^2$, as required.]
There is also a positive contribution when the $a$ and $b$ triplets are paired 
$(aa)(bb)$ in two different loops, but none when they are paired $(ab)(ab)$.
Hence,
\begin{multline} \label{Ma2Mb2count}
\bigl[ (M^a)^2(M^b)^2\bigr]_{\mathcal{L}} = N^2 + 2N\sum_{\alpha}k_\alpha(k_\alpha-1)\\
+ \sum_\alpha \Bigl(\frac{2}{3}k_\alpha^4 - 2k_\alpha^3 + \frac{4}{3}k_\alpha^2  \Bigr)
- \sum_\alpha \Bigl(\frac{1}{3}k_\alpha^4-\frac{1}{3}k_\alpha^2 \Bigr)\\
+ \sum_{\alpha\neq\beta}k_\alpha(k_\alpha-1)k_\beta(k_\beta-1),
\end{multline}
which simplifies to
\begin{equation} \label{EQ:Ma2Mb2}
\bigl[ (M^a)^2(M^b)^2\bigr]_{\mathcal{L}}
= \sum_\alpha \Bigl(\frac{2}{3}k_\alpha^2 - \frac{2}{3}k_\alpha^4\Bigr)
+ \biggl(\sum_{\alpha} k_\alpha^2\biggr)^2.
\end{equation}

It is important to verify that these expressions are compatible with those of
Sects.~\ref{SEC:Correlation} and \ref{SEC:Cumulant}.
First, spin isotropy demands that 
$\langle (\hat{M}^a)^2 \rangle = \tfrac{1}{3}\langle \hat{\mathbf{M}}^2 \rangle$,
which is confirmed by comparing Eqs.~\eqref{EQ:M2kalpha} and \eqref{EQ:MaMb}.
Second, the identity
\begin{equation}
\hat{\mathbf{M}}^4
= \sum_a (\hat{M}^a)^4 + \sum_{a\neq b} (\hat{M}^a)^2(\hat{M}^b)^2
\end{equation}
requires that 3~$\times$~Eq.~\eqref{EQ:Ma4} and 6~$\times$~Eq.~\eqref{EQ:Ma2Mb2}
sum to Eq.~\eqref{EQ:M4kalpha}, which is indeed true.
In summary, the loops estimators for components of the staggered magnetization
at second and fourth order are
\begin{align}
(M^aM^b)_{\mathcal{L}} &= \delta^{ab}\sum_{\alpha} k^2_{\alpha}, \\
\begin{split}
\bigl[ (M^a)^2(M^b)^2 \bigr]_{\mathcal{L}}  &= \frac{2}{3}\sum_{\alpha} \bigl[
\bigl(1-\delta^{ab}\bigr)k^2_{\alpha}-\bigl(1+2\delta^{ab}\bigr)k^4_{\alpha}\bigr)\bigr]\\
&\quad +\bigl(1+2\delta^{ab}\bigr)\biggl(\sum_{\alpha} k^2_{\alpha}\biggr)^2.
\end{split}
\end{align}

\section{\label{SEC:Neel}The N\'{e}el State}

For a bipartite spin system, in which the $A$ and $B$ site labels designate
genuine sublattices, the N\'{e}el state can be written in terms of any AB valence 
bond configuration. This freedom exists because the staggered moments can be 
organized into spin-up/spin-down pairs such that
\begin{equation}
\lvert R \rangle = \prod_{i \in A}b^\dagger_{i\uparrow}\prod_{j \in B}b^\dagger_{j\downarrow}\lvert
\text{vac} \rangle = \prod_{n=1}^N b^\dagger_{i_n\uparrow}b^\dagger_{j_{Pn}\downarrow}\lvert
\text{vac} \rangle
\end{equation}
for any $P \in \mathcal{S}_N$. Since
$b^\dagger_{i\uparrow}b^\dagger_{j\downarrow} = \frac{1}{\sqrt{2}}(\chi^{0\dagger}_{ij}
+ i\chi^{3\dagger}_{ij})$, the N\'{e}el state can be viewed as a superposition of
states with a fixed (but arbitrary) bond configuration and total triplet number ranging from $0$ to $N$:
\begin{equation} \label{EQ:Neelstate}
\lvert R \rangle = \frac{1}{2^{N/2}} \Bigl[ \lvert P \rangle
+ i \sum_{n=1}^N \lvert P; 3_n \rangle - \sum_{m<n}\lvert P; 3_m,3_n \rangle 
+ \cdots \Bigr].
\end{equation}
If we define $t_a(\vec{\mu}) = \sum_{n=1}^N \delta_{\mu_n,a}$, which counts the number of
$a$-triplets in the index vector $\vec{\mu}$, then the overlap of $\lvert R \rangle$ 
with an extended valence bond state $\lvert P; \vec{\mu} \rangle$ can be written as
\begin{equation} \label{EQ:Neeloverlap}
\langle R | P; \vec{\mu} \rangle = \delta_{t_1,0}\delta_{t_2,0}
e^{i(\pi/2) t_3} 2^{-N/2}.
\end{equation}
The overlap vanishes if the state $\lvert P; \vec{\mu} \rangle$ contains any $1$- or $2$-triplets.
Otherwise it is independent of $P$ and constant up to an overall phase factor with an
angle given by $\pi/2 \times \text{the number of $3$-triplets}$.
The derivation of Eq.~\eqref{EQ:Neeloverlap} relies on nothing more than the 
(species) orthogonality of bonds that connect the same two sites
[expressed in Eq.~\eqref{EQ:chi2-commutator}]. There is no need to account
for bond-configurational mismatch because the permutation $P$ in Eq.~\eqref{EQ:Neelstate} 
can be chosen equal to the permutation in $\lvert P; \vec{\mu} \rangle$.

Matrix elements taken between the N\'{e}el state and a valence bond singlet state are particularly
easy to compute. For example, applying Eq.~\eqref{EQ:gammainjnP}, we find that
\begin{equation}
\frac{\langle R \rvert \hat{\gamma}_{i_nj_m} \lvert P \rangle}{\langle R | P \rangle}
= \biggl(\frac{1}{2}\biggr)^{1-\delta_{Pn,m}}
\frac{\langle R | (Pn\ m) P \rangle}{\langle R | P \rangle},
\end{equation}
where the ratio of overlaps on the right-hand side is 1 since both 
$\langle R | P \rangle$ and $\langle R | (Pn\ m) P \rangle$ have the value $2^{-N/2}$.
In general, the matrix element picks up a factor of $\tfrac{1}{2}$ for each offdiagonal
$\hat{\gamma}$ operation (keeping in mind that subsequent operations act on the
reconfigured bonds). The evaluation rule can be expressed as
\begin{equation}
\frac{\langle R \rvert \overbrace{\hat{\gamma}_{ij}\hat{\gamma}_{kl}\cdots}^{\text{$n$ operators}} \lvert P \rangle}{\langle R | P \rangle}
= 2^{N_{\circlearrowleft}-n},
\end{equation}
where $N_{\circlearrowleft}$ now refers to the number of closed loops that are formed
by the bonds and $\hat{\gamma}$ vertices.

\begin{figure}
\begin{center}
\includegraphics{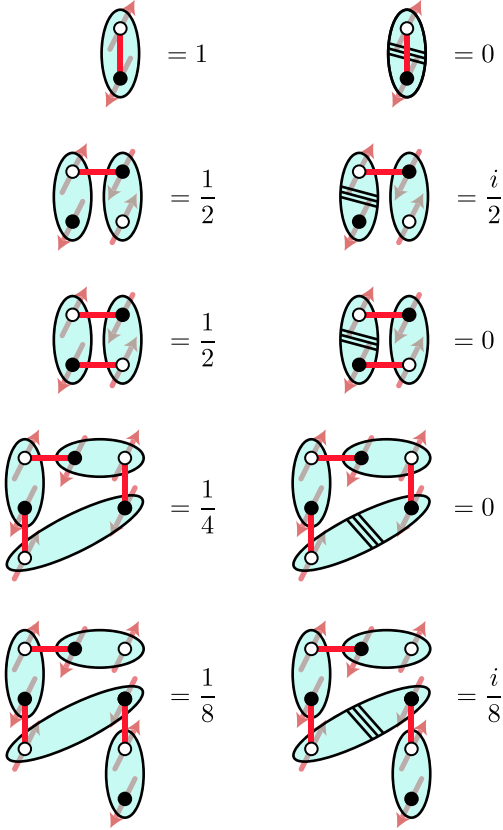}
\end{center}
\caption{\label{FIG:referenve-overlap}
Matrix elements of $\hat{\gamma}$ taken between a valence bond state 
and the reference N\'{e}el state. Closed loops containing one triplet
do not contribute, whereas open strings with one triplet contribute 
the same value as their singlet-bond-only counterpart.}
\end{figure}

In analogy with Eq.~\eqref{EQ:QvuPmuoverlap}, the rules can be extended to cover
states with arbitrary numbers of triplets: 
\begin{equation}
\frac{\langle R \rvert \overbrace{\hat{\gamma}_{ij}\hat{\gamma}_{kl}\cdots}^{\text{$n$ operators}} \lvert P; \vec{\mu} \rangle}{\langle R | P \rangle}
= 2^{N_{\circlearrowleft}-n} \prod_{\alpha=1}^{N_{\circlearrowleft}+N_{\looparrowleft}} K_\alpha.
\end{equation}
In this case, the index $\alpha$ ranges over both the closed loops and the open strings ($N_{\looparrowleft}$
in number) that are formed. 
A closed loop involving $k \le n$ operators and bonds numbered $1,2,  \ldots, k$ has associated with it
a factor
\begin{equation}
K_{\alpha} = \frac{1}{2}\tr \bigl( \tau^{0\dagger}\tau^{\mu_1}\tau^{0\dagger}\tau^{\mu_2}
\cdots \tau^{0\dagger}\tau^{\mu_k}\bigr).
\end{equation}
An open string involving $k \le n$ operators and bonds numbered $1,2,\ldots,k+1$ has a factor
\begin{equation}
K_{\alpha} = \frac{i}{2}\tr \bigl( \tau^{3\dagger}\tau^{\mu_1}\tau^{0\dagger}\tau^{\mu_2}
\cdots \tau^{0\dagger}\tau^{\mu_{k+1}}\bigr).
\end{equation}
Note that any lone triplet in a closed loop gives $K_{\alpha}=0$, whereas 
a lone 3-triplet in an open string gives $K_{\alpha}=i$. See Fig.~\ref{FIG:referenve-overlap}.

\section{\label{SEC:Singlet-triplet}Singlet-triplet Gap}

Consider a quantum spin system whose hamiltonian $\hat{H}$ has a
global singlet ground state 
$\lvert \psi_0 \rangle$.
Starting from an arbitrary singlet trial state 
\begin{equation}
\lvert \psi^{\text{trial}}_{0} \rangle = \sum_P c_P^{\text{trial}} \lvert P \rangle,
\end{equation}
the true ground state wavefunction can be obtained by projection:
\begin{equation} \label{EQ:projtrial0}
\lvert \psi_{0} \rangle = \lim_{\tau\to\infty} e^{-\tau\hat{H}} \lvert \psi^{\text{trial}}_{0} \rangle
= \sum_P c_P \lvert P \rangle.
\end{equation}
From the eigen-equation $\hat{H}\lvert \psi_{0} \rangle = E_0\lvert \psi_{0} \rangle$,
the ground state energy can be isolated by acting from the left with an appropriate reference state:
\begin{equation}
E_0 = \frac{\langle R \rvert \hat{H} \lvert \psi_{0}\rangle}{\langle R | \psi_{0}\rangle}.
\end{equation}
For our purposes, it is most convenient to use a N\'{e}el reference state $\lvert R \rangle$, since
it has a constant-magnitude overlap with all states in the full Hilbert space.
Thus, 
\begin{equation} \label{EQ:E0_W}
E_0 = \frac{\langle R \rvert \hat{H} \lvert \psi_{0}\rangle}{\langle R | \psi_{0}\rangle}
= \frac{ \sum_P W(P) \frac{\langle R \rvert \hat{H} \lvert P\rangle}{\langle R | P\rangle}}{\sum_P W(P)}
\end{equation}
with $W(P) = c_P \langle R | P \rangle = 2^{-N/2} c_P$.
This formulation is similar to that of Eq.~\eqref{EQ:psiOpsi}, except that here
the basic objects of interest are bonds rather than loops.
We can view the ground state energy $E_0 = \langle H_\mathcal{B} \rangle_W$
as the expectation value of the bond estimator for the hamiltonian, $H_\mathcal{B}$, 
in a fluctuating gas of valence bonds.

\begin{figure}
\begin{center}
\includegraphics{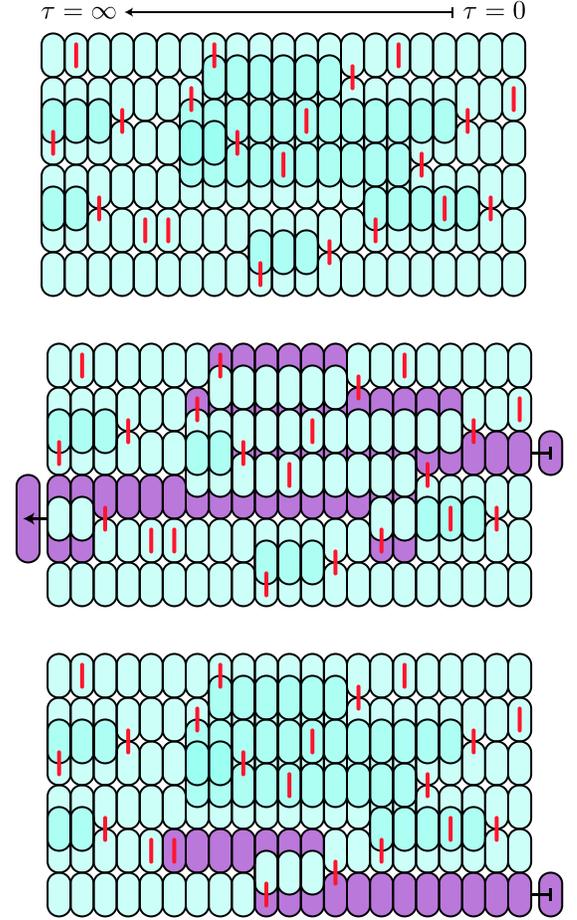}
\end{center}
\caption{\label{FIG:gauntlet} (Top) The evolution operator (in series expansion)
involves long series of
$-\hat{H}_{ij}$ operator strings that map VB configurations in the trial
state to VB configurations in the true ground state through
sequential reordering of the bonds. In the figure, the propagation index~\cite{Sandvik97}
increases from right to left along the horizontal axis. A set of singlet valence
bonds are arranged along the vertical spatial axis.
The red bars denote nearest neighbour interactions.
(Middle) The same operator string can be reinterpreted as acting on a trial
state with one triplet, marked here in purple. 
(Bottom) A triplet state that is acted on directly by a diagonal operation is annihilated.
 }
\end{figure}

Similarly, by constructing a trial wavefunction in the triplet sector,
\begin{equation}
\lvert \psi^{\text{trial}}_{1}\rangle = \hat{M}^3 \lvert \psi^{\text{trial}}_{0} \rangle = i\sum_{n=1}^N \sum_P c_P^{\text{trial}} \lvert P; 3_n \rangle,
\end{equation}
we can project onto the lowest-energy triplet state:
\begin{equation}
\lvert \psi_{1} \rangle = \lim_{\tau\to\infty} e^{-\tau\hat{H}} \hat{M}^3\lvert \psi^{\text{trial}}_{0} \rangle.
\end{equation}
The structure of this projection is closely related to that of Eq.~\eqref{EQ:projtrial0}. In the singlet case,
the evolution operator consists of a long string of $-\hat{H}_{ij}$ operators that successively
reconfigure the bond configuration $\lvert P \rangle$ of the trial state.
Here, we simply need to compute the result of the same strings operating
on $\lvert P; 3_n \rangle$. We know, however, that the update rules for singlet
and triplet bonds are identical except that a direct diagonal operation kills a triplet.
It is straightforward to reinterpret the operator string by designating one
bond of the trial state as a triplet and tracing its evolution. 
As shown in Fig.~\ref{FIG:gauntlet}, the triplet will either run the gauntlet or die in the attempt.
Accordingly, we can write
\begin{equation}
\lvert \psi_{1} \rangle = i \sum_{n=1}^N \sum_P  g_n(P)\,c_P \lvert P; 3_n\rangle,
\end{equation}
where $g_n(P)$ counts the number of surviving triplets whose final destination
is the $n^{\text{th}}$ bond of configuration $P$. By definition,
\begin{equation}
0 \le \sum_{n=1}^N g_n(P) \le N.
\end{equation}

The energy of the low-lying triplet state is
\begin{equation} 
E_1 = \frac{\langle R \rvert \hat{H} \lvert \psi_{1}\rangle}{\langle R | \psi_{1}\rangle}
= \frac{ \sum_P W(P) \sum_{n=1}^Ng_n(P) \frac{\langle R \rvert \hat{H} \lvert P; 3_n\rangle}{\langle R | P;3_n\rangle}}{\sum_P W(P) \sum_{n=1}^Ng_n(P)}.
\end{equation}
Here we have made use of the fact that $\langle R | P; 3_n \rangle = i\langle R | P \rangle$.
Alternatively, $E_1$ can be expressed as
\begin{equation} \label{EQ:E1_W}
E_1 =   \frac{\bigl\langle \sum_n g_n H_{\mathcal{B}} \bigr\rangle_W}
{\bigl\langle \sum_n g_n \bigr\rangle_W}
\end{equation}
since $\langle R \rvert \hat{H} \lvert P; 3_n\rangle/\langle R | P;3_n\rangle$
and $\langle R \rvert \hat{H} \lvert P \rangle/\langle R | P \rangle$ differ
only when $g_n = 0$.
Combining Eqs.~\eqref{EQ:E0_W} and \eqref{EQ:E1_W} gives
\begin{equation} \label{EQ:gap}
E_1 - E_0 = \frac{\bigl\langle \sum_n g_n H_{\mathcal{B}} \bigr\rangle_W
-\bigl\langle \sum_n g_n \bigr\rangle_W \langle H_{\mathcal{B}} \rangle_W}
{\bigl\langle \sum_n g_n \bigr\rangle_W}.
\end{equation}

A numerical measurement of the singlet-triplet gap via Eq.~\eqref{EQ:gap}
has the advantage that it can be carried out during a simulation of the
$S=0$ ground state properties. This gives a substantial error cancellation
in stochastic methods over techniques where $E_0$ and $E_1$ are 
computed individually and subtracted. In some systems, however, the
lowest-energy singlet and triplet states may differ so substantially that
the reweighting of the operator string becomes inefficient. In that case,
so few triplets survive the projection that $\langle \sum_n g_n \rangle \approx 0$.

\section{\label{SEC:Stiffness} Spin stiffness }

Again, let us imagine that the $S=0$ ground state is obtained via projection:
\begin{equation}
\lvert \psi \rangle = \lim_{\tau\to\infty} e^{-\tau\hat{H}} \lvert \psi^{\text{trial}} \rangle.
\end{equation}
Now suppose that we introduce a twist field $\phi(\mathbf{r})$ 
representing a local rotation of the spins about the 3 axis.
We will be concerned with the limit in which the field gradient 
$\nabla \phi(\mathbf{r})$ is small with respect to the lattice spacing.
For concreteness, we endow the field with a single long-wavelength mode
$\phi(\mathbf{r}) = \phi_0 + \mathbf{r}\cdot\mathbf{q}$. As a consequence, 
the relative spin rotation angles satisfy $\theta_{ij} = (\mathbf{r}_i - \mathbf{r}_j)\cdot\mathbf{q}$.

\begin{figure*}
\includegraphics{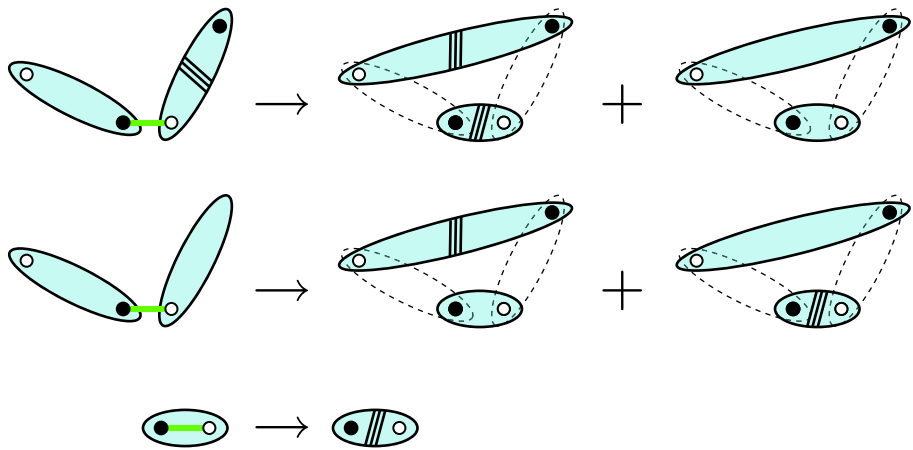}
\caption{ \label{FIG:updates3}
Several possible bond reconfigurations are shown, corresponding
to action by each term in $\hat{T}^a$. The relevant prefactors are suppressed.}
\end{figure*}

Equation~\eqref{EQ:chidaggerchirotate} then suggests that the hamiltonian---assuming it has the form 
of Eq.~\eqref{EQ:modelH}---transforms as
\begin{equation} \label{EQ:H-twisted}
-\hat{H} \to -\hat{H}[\mathbf{q}] = -\hat{H} + q^a\hat{T}^a + \frac{1}{2}q^aq^b\hat{G}^{ab},
\end{equation}
where the gradient and Hessian are given by
\begin{equation}\label{EQ:Ta}
\hat{T}^a = \frac{1}{2}\sum_{\langle ij \rangle}J_{ij} (\mathbf{r}_i-\mathbf{r}_j)\cdot\mathbf{e}^a
\Bigl(\chi^{0\dagger}_{ij}\chi^{3}_{ij}+\chi^{3\dagger}_{ij}\chi^{0}_{ij}\Bigr) + \cdots
\end{equation}
and
\begin{multline} \label{EQ:Gab}
\hat{G}^{ab} = \frac{1}{2}\sum_{\langle ij \rangle}J_{ij} (\mathbf{r}_i-\mathbf{r}_j)\cdot\mathbf{e}^a
(\mathbf{r}_i-\mathbf{r}_j)\cdot\mathbf{e}^b\\
\times\Bigl(-\chi^{0\dagger}_{ij}\chi^{0}_{ij}
+ \chi^{3\dagger}_{ij}\chi^{3}_{ij}\Bigr) + \cdots
\end{multline}
Here, $+ \cdots$ represents additional interaction terms in $K_{ijkl}$ and beyond.
Equation~\eqref{EQ:Ta} is a spin current operator that changes the number of 
triplet bonds by $\pm1$. (See Fig.~\ref{FIG:updates3}.) It does so in proportion to the 
interaction strength and the projection of the bond length onto the axis of propagation.
Equation~\eqref{EQ:Gab}, on the other hand, changes the triplet count by 
$0$ or $\pm 2$.

Under the influence of the twist field, the ground state energy is
\begin{equation}
E[\mathbf{q}] = \lim_{\tau\to\infty} \frac{ \langle R \lvert \hat{H}[\mathbf{q}] e^{-\tau \hat{H}[\mathbf{q}]} \rvert \psi_0 \rangle }{ \langle R \lvert e^{-\tau \hat{H}[\mathbf{q}]} \rvert \psi_0 \rangle }.
\end{equation}
By virtue of the variational principle, $E[\mathbf{q}]$ must be stationary
with respect to variation.
Its convexity at $\mathbf{q}=0$ is a measure of the spin stiffness.
Derivatives $\partial_a =  \partial/\partial q^a$ can be handled
using the identity
\begin{equation}
\int_0^\tau \! d\tau' \, e^{-(\tau-\tau')\hat{H}} \partial_a \hat{H}
e^{-\tau'\hat{H}}
= - \partial_a\bigl( e^{-\tau\hat{H}} \bigr).
\end{equation}
Hence,
\begin{equation} \label{EQ:Eqdqa}
\frac{\partial E[\mathbf{q}]}{\partial q^a} \biggr\rvert_{\mathbf{q}=0}
= \frac{\langle R \lvert \hat{T}^a \rvert \psi \rangle}
{\langle R | \psi \rangle}
+ \frac{\langle R \lvert (\hat{H}-E) \rvert \psi^a \rangle}
{\langle R | \psi \rangle} = 0,
\end{equation}
with $\lvert \psi^a \rangle = \partial_a \lvert \psi \rangle\bigr\rvert_{\mathbf{q}=0}$
given by
\begin{equation}
\lvert \psi^a \rangle =
\int_0^\tau \! d\tau' \, e^{-(\tau-\tau')\hat{H}} \hat{T}^a e^{-\tau'\hat{H}}\lvert \psi^{\text{trial}} \rangle.
\end{equation}
The state $\lvert \psi^a \rangle$ evolves from the
singlet trial state but differs from the projected
ground state in that it is constructed with an $\mathbf{e}^a$-directed triplet injected
at all times $\tau' < \tau$.

Making use of Eq.~\eqref{EQ:Eqdqa}, we can 
show that the spin stiffness is given by
\begin{widetext}
\begin{equation} \label{EQ:spinstiffness}
\rho^{ab}_{\text{s}} = \frac{\partial^2 E[\mathbf{q}]}{\partial q^a\partial q^b}
\biggr\rvert_{\mathbf{q}=0}
= \frac{\langle R \lvert \hat{G}^{ab} \rvert \psi \rangle}
{\langle R | \psi \rangle}
+ \frac{\langle R \lvert \hat{T}^{a} \rvert \psi^b \rangle}
{\langle R | \psi \rangle}
+ \frac{\langle R \lvert \hat{T}^{b} \rvert \psi^a \rangle}
{\langle R | \psi \rangle}
+ \frac{\langle R \lvert (\hat{H}-E) \rvert \psi^{ab} \rangle}
{\langle R | \psi \rangle},
\end{equation}
where
\begin{equation}
\begin{split}
\lvert \psi^{ab} \rangle &= 
\int_0^\tau \! d\tau' \! \int_{\tau'}^\tau \! d\tau'' \, e^{-(\tau-\tau'')\hat{H}} \hat{T}^a e^{-(\tau''-\tau')\hat{H}} \hat{T}^b e^{-\tau'\hat{H}}\lvert \psi^{\text{trial}} \rangle \\
&+\int_0^\tau \! d\tau' \! \int_{\tau'}^\tau \! d\tau'' \, e^{-(\tau-\tau'')\hat{H}} \hat{T}^b e^{-(\tau''-\tau')\hat{H}} \hat{T}^a e^{-\tau'\hat{H}}\lvert \psi^{\text{trial}} \rangle \\
& + \int_0^\tau \! d\tau' \, e^{-(\tau-\tau')\hat{H}} \hat{G}^{ab} e^{-\tau'\hat{H}}\lvert \psi^{\text{trial}} \rangle.
\end{split}
\end{equation}
\end{widetext}
The state $\lvert \psi^{ab} \rangle$ evolves from the
singlet trial state $\lvert \psi^{\text{trial}} \rangle$ with two $\mathbf{e}^a$- and
$\mathbf{e}^b$-directed triplets injected at times $\tau'$ and $\tau''$.

Equation~\eqref{EQ:spinstiffness} can be evaluated using the reweighting
trick introduced in Sect.~\ref{SEC:Singlet-triplet}. 
The same operator sequence used to generate the singlet ground state can
be reinterpreted to include all possible triplet pairs (at all possible
starting locations, not simply at the $\tau = 0$ end as was the case
for the singlet-triplet gap measurement). The spin stiffness is related
to the fluctuations in energy of the final bond configuration arising
from operator sequences in which neither triplet is annihilated.

\section{Summary}

We have presented in detail a formal framework for organizing calculations in the
valence bond basis. This approach is based on manipulations of valence bond 
creation and annihilation operators, $\chi^{\mu\dagger}_{ij}$ and $\chi^{\mu}_{ij}$,
that act between any two sites of the spin lattice. 
These operators are similar to those introduced in Ref.~\onlinecite{Sachdev90} for fixed dimer configurations but are endowed with an expanded operator algebra that is
compatible with the overcomplete basis of all possible dimer configurations.

We have focused on what we call the AB valence bond basis, in 
which half the lattice sites are assigned the label A and the other half B
and only dimers connecting sites with different labels are allowed.
(The use of AB bonds only is routine in the literature, especially when
the hamiltonian of interest does not contain explicitly frustrating AA/BB interactions.)
The states in this restricted basis are uniquely characterized by 
permutations of the B site labels. Overlaps and matrix elements of such
states are related in a systematic way to the cycle structure of the permutations.
We have shown, for example, that the presence of antiferromagnetism in a bipartite system
is related to the long tail behaviour of the cycle length distribution $\langle n_k \rangle_W$.

Correlation functions of the isotropic spin interaction 
$\mathbf{S}_i \cdot \mathbf{S}_j$ are related 
to the cumulants of the operator
$\hat{\gamma}_{ij} = \frac{1}{4}\bigl(1+\epsilon_{ij}\bigr) 
- \epsilon_{ij} \chi^{0\dagger}_{ij} \chi^{0}_{ij}$.
The various contributions can thus be computed in a straightforward way
from the set of connected Goldstone diagrams---albeit with the diagrams
interpreted quite differently than they are in the usual quantum many-body physics context.
(Moreover, there is no need to worry about which of $i$ and $j$ are A or B sites.
The different cases are all handled automatically by the sign factor $\epsilon_{ij}$.)
As an example of this approach, we have derived explicit expressions for second-, 
fourth-, and sixth-order expectation values of the staggered magnetization.

We have also emphasized that, when triplet bonds are fully accounted for,
the valence bond basis spans the entire Hilbert space, not merely the $S=0$ subspace.
The valence bond operators carry an index $\mu$ that specifies the singlet ($\mu = 0$) or 
triplet ($\mu = 1,2,3$) nature of each bond. We have extended the rules for computing
overlaps and matrix elements to include states with triplet bonds. 
As an example of how this can be useful, we have derived
expressions for the singet-triplet gap and the spin stiffness
that can be interpreted a reweightings of the
operator string in a projected singlet ground state.
The reweighting formulation is ideally suited for use in valence bond projector Monte Carlo.

This material is based upon work supported by the National Science 
Foundation under Grant No.\ DMR-0513930.

\appendix

\section{\label{SEC:Permutation}}
A permutation $P \in \mathcal{S}_N$ is a bijective map that takes $n$ to $Pn$
(and $\bar{P}n$ to $n$) for every $1 \le n \le N$:
\begin{equation} \label{EQ:permPexplicit}
P = \begin{pmatrix}
1 & P1\\
2 & P2\\
\vdots & \vdots \\
N & PN
\end{pmatrix}.
\end{equation}
An alternative notation for Eq.~\eqref{EQ:permPexplicit}
involves cycles of the form $(1\ 2\ 3)$, which expresses
the functional relationship $1\to2$, $2\to 3$, and $3 \to 1$.
Every permutation can be decomposed into a product of disjoint cycles
\begin{equation} \label{EQ:cycledecomp}
P = \prod_{\alpha=1}^{\cycles(P)}
( \xi_\alpha\ P \xi_\alpha\ P^2 \xi_\alpha\ \cdots \ P^{k_\alpha-1} \xi_\alpha ),
\end{equation}
where $\cycles(P)$ is the number of cycles in $P$, $ \xi_\alpha$ is 
a representative number (the lowest, say) in a particular cycle, and $k_\alpha$ is
its length. Two numbers $n$ and $m$ are in the same cycle of $P$ iff
there is some integer power $k \ge 0$ such that $P^kn = m$, which we denote
$n \overset{P}{\sim} m$. It follows that $n$ is in the cycle labelled $\alpha$
iff $n \overset{P}{\sim} \xi_{\alpha}$.

The cycle decomposition given in Eq.~\eqref{EQ:cycledecomp} is unique
and requires the fewest possible number of cycles. One can, however,
further decompose each term into several smaller cycles that are no longer disjoint
(\emph{i.e.}, they may have elements in common).
This process can always be carried further until the permutation
is expressed entirely as a product of transpositions (cycles of length 2).
Accordingly, to understand how the cycle structure of $\bar{Q}P'$ differs from that
of $\bar{Q}P$, we need only consider the sequence of changes
induced by the chain of transpositions that separates $P'$ from $P$.

The result of a single transposition acting on $P$ is
\begin{align}
(n\ m)P &= \begin{pmatrix}
1 & P1\\
2 & P2\\
\vdots & \vdots \\
\bar{P}n & m \\
\vdots & \vdots \\
\bar{P}m & n \\
\vdots & \vdots \\
N & PN
\end{pmatrix}.
\intertext{Hence,}
\bar{Q}(n\ m)P &= \begin{pmatrix}
1 & \bar{Q}P1\\
2 & \bar{Q}P2\\
\vdots & \vdots \\
\bar{P}n & \bar{Q}m \\
\vdots & \vdots \\
\bar{P}m & \bar{Q}n \\
\vdots & \vdots \\
N & P\bar{Q}N
\end{pmatrix}.
\end{align}
As to how this differs from $\bar{Q}P$, there are three possibilities to consider.
First, when $n=m$, the 2-cycle is just an identity element; the equalities
$\bar{Q}P = \bar{Q}(n\ m)P$ and $\cycles(\bar{Q}P) = \cycles(\bar{Q}(n\ m)P)$
follow trivially.
Otherwise, the outcome depends on whether $\bar{P}n$
and $\bar{P}m$ are in the same cycle. If they are,
the transposition splits one cycle into two,
\begin{align}
\bar{Q}P &= (1\ 8\  3\ \bar{P}n\ 2\ 7\ 4\ \bar{P}m\ 6\ 5)\cdots \\
\bar{Q}(n\ m)P &= (1\ 8\ 3\ \bar{P}m\ 6\ 5)(2\ 7\ 4\ \bar{P}n)\cdots,
\end{align}
and $\cycles(\bar{Q}(n\ m)P) - \cycles(\bar{Q}P) = +1$.
If $\bar{P}n$ and $\bar{P}m$ are in different cycles,
the transposition merges two cycles into one,
\begin{align}
\bar{Q}P &= (1\ 3\ 7\ \bar{P}n\ 4\ 6)(2\ 8\ \bar{P}m\ 5)\cdots \\
\bar{Q}(n\ m)P &= (1\ 3\ 7\ \bar{P}m\ 5\ 2\ 8\ \bar{P}m\ 4\ 6)\cdots,
\end{align}
and $\cycles(\bar{Q}(n\ m)P) - \cycles(\bar{Q}P) = -1$.
These three cases can be summarized by
\begin{equation}
\cycles(\bar{Q}(n\ m)P) - \cycles(\bar{Q}P) + \delta_{n,m} 
= \begin{cases}
+1 & \text{if $\bar{P}n \overset{\bar{Q}P}{\sim} \bar{P}m$ } \\
-1 & \text{otherwise}
\end{cases}
\end{equation}
or, equivalently,
\begin{equation}
\frac{\langle Q \lvert (n\ m) P\rangle}{\langle Q | P \rangle} \biggl(\frac{1}{2}
\biggr)^{1-\delta_{n,m}} 
= \begin{cases}
1 & \text{if $\bar{P}n \overset{\bar{Q}P}{\sim} \bar{P}m$ } \\
\frac{1}{4} & \text{otherwise}.
\end{cases}
\end{equation}

\end{document}